\documentclass[twocolumn]{aastex63}
\usepackage{amsmath}

\submitjournal{ApJ}
\shorttitle{The Bimodal Origins of Early-Type Stars Magnetism}
\shortauthors{Jermyn and Cantiello}

\definecolor{purple}{rgb}{0.5,0,0.5}
\definecolor{darkgreen}{rgb}{0.1,0.6,0.1}
\definecolor{orange}{rgb}{1,0.6,0}

\newcommand{\brvs}{Brunt-V\"ais\"al\"a}

\begin{document}

\title{The Origin of the Bimodal Distribution of Magnetic Fields in Early-type Stars}

\correspondingauthor{Adam S. Jermyn}
\email{adamjermyn@gmail.com}

\author[0000-0001-5048-9973]{Adam S. Jermyn}
\affiliation{Center for Computational Astrophysics, Flatiron Institute, New York, NY 10010, USA}

\author[0000-0002-8171-8596]{Matteo Cantiello}
\affiliation{Center for Computational Astrophysics, Flatiron Institute, New York, NY 10010, USA}
\affiliation{Department of Astrophysical Sciences, Princeton University, Princeton, NJ 08544, USA}

\begin{abstract}

In early-type stars a fossil magnetic field may be generated during the star formation process or be the result of a stellar merger event.
Surface magnetic fields are thought to be erased by (sub)surface convection layers, which typically leave behind weak disordered fields.
However, if the fossil field is strong enough it can prevent the onset of (sub)surface convection and so be preserved onto the main sequence.
We calculate the critical field strength at which this occurs, and find that it corresponds well with the lower limit amplitude of observed fields in strongly magnetised Ap/Bp stars ($\approx$ 300 G). The critical field strength is predicted to increase slightly during the main sequence evolution, which could also explain the observed decline in the fraction of magnetic stars.
This supports the conclusion that the bimodal distribution of observed magnetic fields in early-type stars reflects two different field origin stories: strongly magnetic fields are fossils fields inherited from star formation or a merger event, and weak fields are the product of on-going dynamo action.

\end{abstract}


\section{Introduction}

Magnetic fields play many roles in stellar evolution.
They are thought to modify stellar winds and enable spin-down~\citep{1967ApJ...148..217W,2009IAUS..259..423U}, transport angular momentum~\citep{2002A&A...381..923S}, and influence accretion~\citep{2007prpl.conf..479B}.
They can enhance chemical mixing~\citep{2019ApJ...870L...5H} or inhibit it~\citep{1966MNRAS.133...85G}.
Moreover, magnetism can influence heat transport, most notably by producing star spots~\citep{Berdyugina2005,2011A&A...534A.140C,2019ApJ...883..106C}.

Despite its importance, the origins of stellar magnetism remains an open question.
Numerical simulations and theoretical arguments demonstrate that dynamo action owing to either differential rotation or convection can amplify infinitesimal seed fields to detectable strength~\citep{2002A&A...381..923S,2009ASPC..416..369B}.
In early-type stars with masses less than $\approx 7 M_\odot$ envelope convection tends to be quite weak \citep{2009A&A...499..279C}, though, so the predicted surface field strengths are of order $1-10\,\mathrm{G}$~\citep{2019ApJ...883..106C}. At higher mass and at solar metallicity, the presence of the iron convection zone (FeCZ) results in stronger dynamo-generated magnetic fields, with surface amplitudes $10-300\,\mathrm{G}$~\citep{2011A&A...534A.140C}.
Simultaneously, the low magnetic diffusivity of stellar matter means that at any strength a stable magnetic field configuration~\citep{2006A&A...450.1077B,2010A&A...517A..58D} can remain frozen in from the early formation of a star through to the main sequence, so long as it is not disturbed by non-diffusive processes~\citep{1945MNRAS.105..166C,2017RSOS....460271B}.
By the same token, stellar mergers are common among massive stars and may be able to generate magnetic fields~\citep{Schneider2019} which can survive through to the main sequence.

Recently, observations have revealed that early-type (A/B/O) stars exhibit a bimodal distribution of surface magnetic field strengths~\citep{2007A&A...475.1053A,2017MNRAS.465.2432G}, and that there is a `desert' range of field strengths in which few or no stars exist~\citep{2015A&A...574A..20F}.
This bimodal distribution raises the possibility of using strength to diagnose the origin of magnetic fields: perhaps strongly magnetized stars have retained fossil fields while weakly magnetized ones are generating their fields via contemporary dynamo processes.

The interaction of a magnetic field with convection has been discussed in the literature~\citep{1966MNRAS.133...85G,1987MNRAS.224.1019M}.
A strong enough magnetic field can suppress convection~\citep{2019MNRAS.487.3904M}, with important consequences for observed stellar properties like macroturbulence~\citep{2013MNRAS.433.2497S}.
On the other hand a slightly weaker large scale magnetic field threading the same convective region can be twisted, losing its large scale and stability properties.

We begin in Section~\ref{sec:criterion} by reviewing the criterion for a magnetic field to suppress convection.
We then present calculations of the critical magnetic field $B_{\rm crit}$ at which subsurface convection is shut off in stars ranging from $2-12 M_\odot$ across the Hertzsprung-Russel (H-R) diagram.
We find that this field is of order $10^2$-$10^4\,\mathrm{G}$.
We then show the strength of the equipartition dynamo field for the same stellar models (Section~\ref{sec:dynamo}), and find that this is generally $10$ to $100$ times smaller.

In Section~\ref{sec:prediction} we combine these results with a simple physical argument for the evolution of the magnetic field in a convective region, and suggest that fossil fields weaker than $B_{\rm crit}$ are erased by convection, while those stronger than $B_{\rm crit}$ are stable.
This naturally produces a bimodal distribution of field strengths as well as the approximate range of field strengths of the magnetic desert.
We compare these and more predictions features with observations in Section~\ref{sec:obs} and find good agreement.
We conclude with a discussion of the astrophysical implications in Section~\ref{sec:discussion}.

\section{Convection Criterion}
\label{sec:criterion}

Magnetic fields make the criterion for convective instability more strict~\citep{1966MNRAS.133...85G}.
This has been studied by multiple authors who gradually incorporated additional effects such as non-ideal gas behavior and radiation pressure~\citep{2009ApJ...700..387M,2019MNRAS.487.3904M}.
The most general stability criterion of which we are aware is~\citep{2019MNRAS.487.3904M}
\begin{align}
\frac{4-3\beta}{\beta}\left(\nabla -\nabla_{\rm ad}\right) - \frac{v_{\rm A,r}^2}{v_{\rm A,r}^2+c_{\rm s}^2}\left(1+\frac{d\ln \Gamma_1}{d\ln p}\right) < 0,
\label{eq:criterion0}
\end{align}
where
\begin{align}
	\nabla \equiv \frac{d\ln T}{d\ln P}
\end{align}
is the temperature gradient in the star,
\begin{align}
	\nabla_{\rm ad} \equiv \left.\frac{\partial\ln T}{\partial\ln P}\right|_s
\end{align}
is the adiabatic temperature gradient,
\begin{align}
	\beta \equiv \frac{p_{\rm gas}}{p}
\end{align}
is the gas pressure fraction,
\begin{align}
	v_{\rm A,r}^2 = \frac{B_r^2}{4\pi \rho}
\end{align}
is the square of the radial Alfv{\'e}n speed, and $\Gamma_1$ is the first adiabatic index, often just called $\Gamma$ owing to its common use.

In thermal equilibrium and in the absence of convection, the temperature gradient equals the radiative temperature gradient
\begin{align}
	\nabla = \nabla_{\rm rad} \equiv \frac{3\kappa L}{64 \pi G M\sigma T^4},
\end{align}
where $\kappa$ is the opacity, $L$ is the luminosity, and $M$ is the mass beneath the point of interest.
Inserting this into equation~\eqref{eq:criterion0} we find
\begin{align}
\frac{4-3\beta}{\beta}\left(\nabla_{\rm rad} -\nabla_{\rm ad}\right) - \frac{v_{\rm A, r}^2}{v_{\rm A, r}^2+c_{\rm s}^2}\left(1+\frac{d\ln \Gamma_1}{d\ln p}\right) < 0.
\label{eq:criterion}
\end{align}
Some algebra then yields the critical radial magnetic field which prevents convection
\begin{align}
	B_{\rm crit, r} = \sqrt{\frac{4 \pi \rho  c_{\rm s}^2  Q \left(\nabla_{\rm rad}-\nabla_{\rm ad}\right)}{1 - Q \left(\nabla_{\rm rad}-\nabla_{\rm ad}\right) + d\ln\Gamma_1/d\ln p}},
	\label{eq:criterion}
\end{align}
where for compactness we have let
\begin{align}
	Q \equiv \frac{4-3\beta}{\beta}.
\end{align}

If the magnetic field is purely horizontal (i.e. $v_{\rm A,r} = 0$) then it does not stabilize linear motions against convection.
However, a mostly horizontal magnetic field only occurs in a small strip at the equator of a dipole field, so we expect that if equation~\eqref{eq:criterion} is satisfied by the overall magnitude of the magnetic field $B$ then it is likely satisfied over most of the solid angle of the star for $B_r$.
Because our arguments in Section~\ref{sec:prediction} are local arguments they are not modified by the existence of a subcritical latitude range, and because the observations in Section~\ref{sec:obs} are of the average field over the surface our interpretation of those is likewise unchanged.
As such we now drop the subscript `r' and just consider the magnitude of the magnetic field, not its geometry.

It is worth noting that equation~\eqref{eq:criterion} is a purely local stability criterion.
There could be scenarios in which the global curvature of the magnetic field weakens this criterion, allowing convection even with larger magnetic fields.
However, because the depth of the convection zone is a small fraction of the radius of the star, of order $1\,\%$, we expect such effects to be small for magnetic fields with large-scale structure.

We evaluated $B_{\rm crit}$ using the Modules for Experiments in Stellar Astrophysics
\citep[MESA][]{Paxton2011, Paxton2013, Paxton2015, Paxton2018, Paxton2019} software instrument.
Details on the microphysics inputs to this software instrument are given in Appendix~\ref{appen:mesa}. The inlists used to run our models can be found on \href{https://doi.org/10.5281/zenodo.3891939}{Zenodo} \citep{data}.

For each stellar model we evaluated the maximum $B_{\rm crit}$ needed to prevent the formation of any subsurface convection zone.
We did this as a function of both mass and evolutionary history for stars ranging from $2-12 M_\odot$, shown in Fig.~\ref{fig:Bcrit}.
Here we present results for an initial metallicity of Z=0.02, but in Appendix~\ref{appen:grids} we report results for model grids with Z = 0.014, 0.006 and 0.002 as well.

At the low-mass end, the critical field is of order $10^{3}\,\mathrm{G}$ and is set by the Helium ionization convection zone (HeCZ). This convection zone is driven by second helium ionization,  we refer to \citet{2019ApJ...883..106C} for a detailed description and classification of envelope convection zones.
With increasing mass the HeCZ becomes weaker and the critical field falls to $5\times 10^2\,\mathrm{G}$.
This can be seen more clearly in Fig.~\ref{fig:HeCZ_Bcrit}, which shows just the field needed to shut off the HeCZ for tracks ranging from $2-6 M_\odot$.

\begin{figure}
\centering
\includegraphics[width=0.47\textwidth]{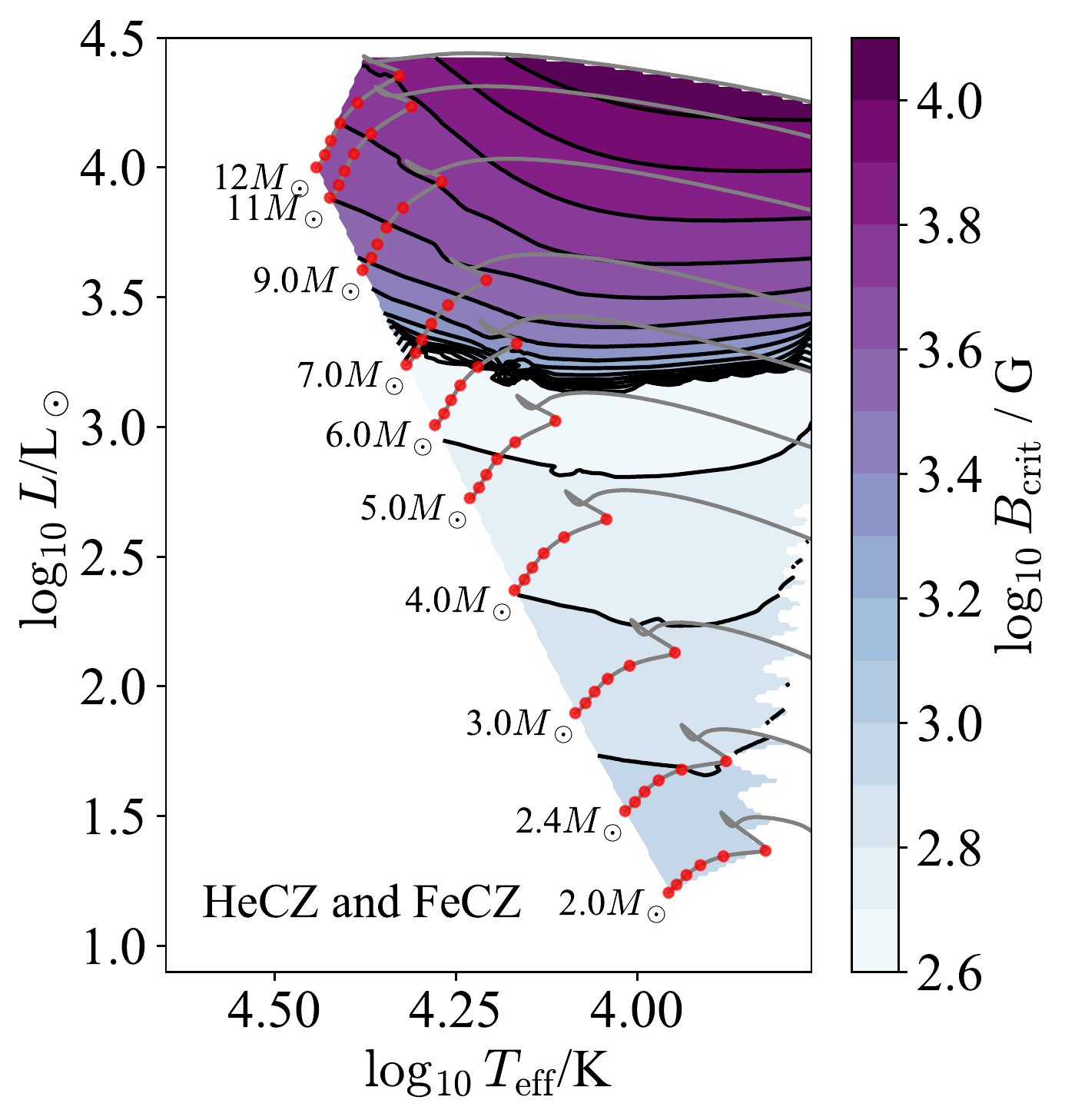}
\caption{The critical magnetic field $B_{\rm crit}$ given by equation~\eqref{eq:criterion} is shown on a Hertzsprung-Russel diagram in terms of $\log T_{\rm eff}$ and $\log L$ for stellar models ranging from $2-12 M_\odot$ with Z=0.02. The sharp increase in $B_{\rm crit}$ at $\log L/L_\odot \approx 3.2$ is due to the appearance of the FeCZ. At low $\log T_{\rm eff}$ and low $\log L$ we have omitted regions where a vigorous H convection merges with the subsurface convection zones, and we do not report values calculated for the H convection zone in this diagram. Red dots show the location of 20\% increase in fractional age, from the zero age main sequence to hydrogen exhaustion.}
\label{fig:Bcrit}
\end{figure}

\begin{figure}
\centering
\includegraphics[width=0.47\textwidth]{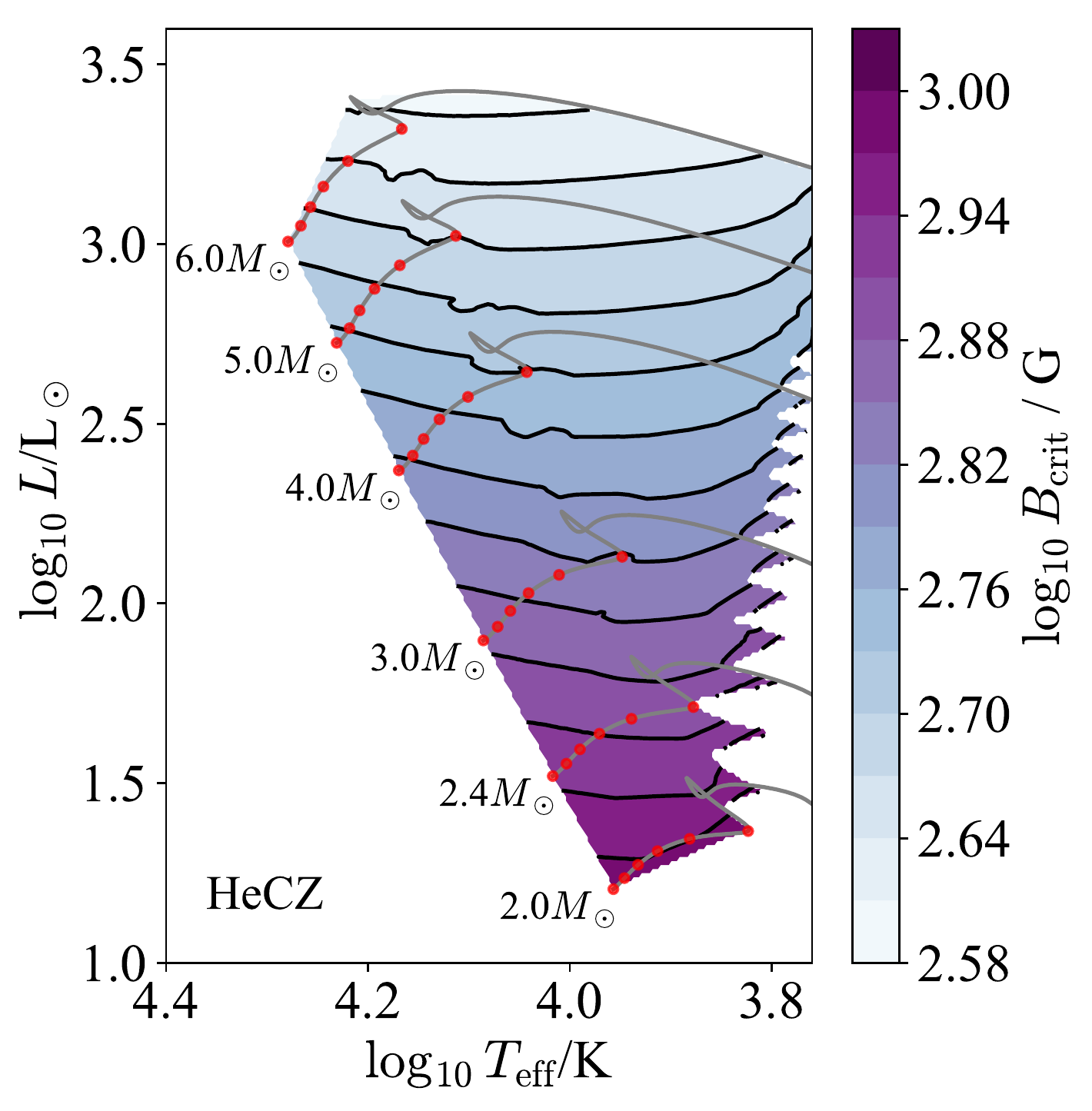}
\caption{The critical magnetic field $B_{\rm crit}$ given by equation~\eqref{eq:criterion} is shown on a Hertzsprung-Russel diagram in terms of $\log T_{\rm eff}$ and $\log L$ for stellar models ranging from $2-6 M_\odot$, which is the range in which the FeCZ is absent and the HeCZ is the most important convection zone.}
\label{fig:HeCZ_Bcrit}
\end{figure}

At higher masses, beginning around $5-7M_\odot$ depending on the age of the star, the FeCZ is much stronger than the HeCZ and so sets $B_{\rm crit}$.
This results in a large jump in $B_{\rm crit}$, to $3\times 10^{3}\,\mathrm{G}$ around a mass of $7M_\odot$, eventually rising to $10^4\,\mathrm{G}$ for $12 M_\odot$ stars.

In both regimes the critical field strength varies over the lifetime of the star.
For stars with $M < 5 M_\odot$ it varies by of order $40$~per-cent on the main sequence, before rapidly declining by a similar amount when crossing the Hertzsprung gap.
For stars with $M > 7 M_\odot$ the field strength first rises by a factor of $2-3$ leading up to the hook, then declines by $50$~per-cent as they cross the Herztsprung gap.
Finally, for stars with $5 M_\odot < M < 7 M_\odot$ the star begins with just the HeCZ, then forms the FeCZ, and in some cases subsequently \emph{loses} the FeCZ, returning to just having the HeCZ.
In this last limit the variation in $B_{\rm crit}$ is most dramatic, with an increase of more than $10$-fold leading up to the hook, followed by a somewhat smaller decrease in the gap.

Because the subsurface convection zones are very inefficient, turning off convection in these layers produces a minimal impact on the radius and $T_{\rm eff}$ of the star.
A comparison of stellar models with subsurface convection versus those without is included in Appendix~\ref{appen:comparison}.

Note that when the models cool below $\log T_{\rm eff}/\mathrm{K} \approx 4$, they develop a surface convection zone driven by hydrogen recombination \citep[e.g.][]{2019ApJ...883..106C}. As a model cools, this convection zone eventually deepens and merges with any existing subsurface convection layer,  dominating the  envelope properties. In this work we did not study these later evolutionary phases, and we restricted our discussion to He and Fe convection zones only.

\section{Dynamo Strength}
\label{sec:dynamo}

The convective dynamo is capable of amplifying small seed fields to significant amplitudes.
The precise field strength at which this saturates is not known, but is believed to be of order the field required to quench convection~\citep{1989ApJ...342.1158M}.
For non-rotating convection zones this results in a field with small-scale structure, coherent over distances of order the convective mixing length, with an approximately equipartition field~\citep{2011IAUS..272...32C}
\begin{align}
	\frac{B^2}{8\pi} \approx \frac{1}{2}\rho v_{\rm c}^2,
\end{align}
or
\begin{align}
	B \approx \sqrt{4\pi \rho v_{\rm c}^2},
	\label{eq:dynamo0}
\end{align}
where $v_{\rm c}$ is the root-mean square of the convective velocity.

\begin{figure}
\centering
\includegraphics[width=0.47\textwidth]{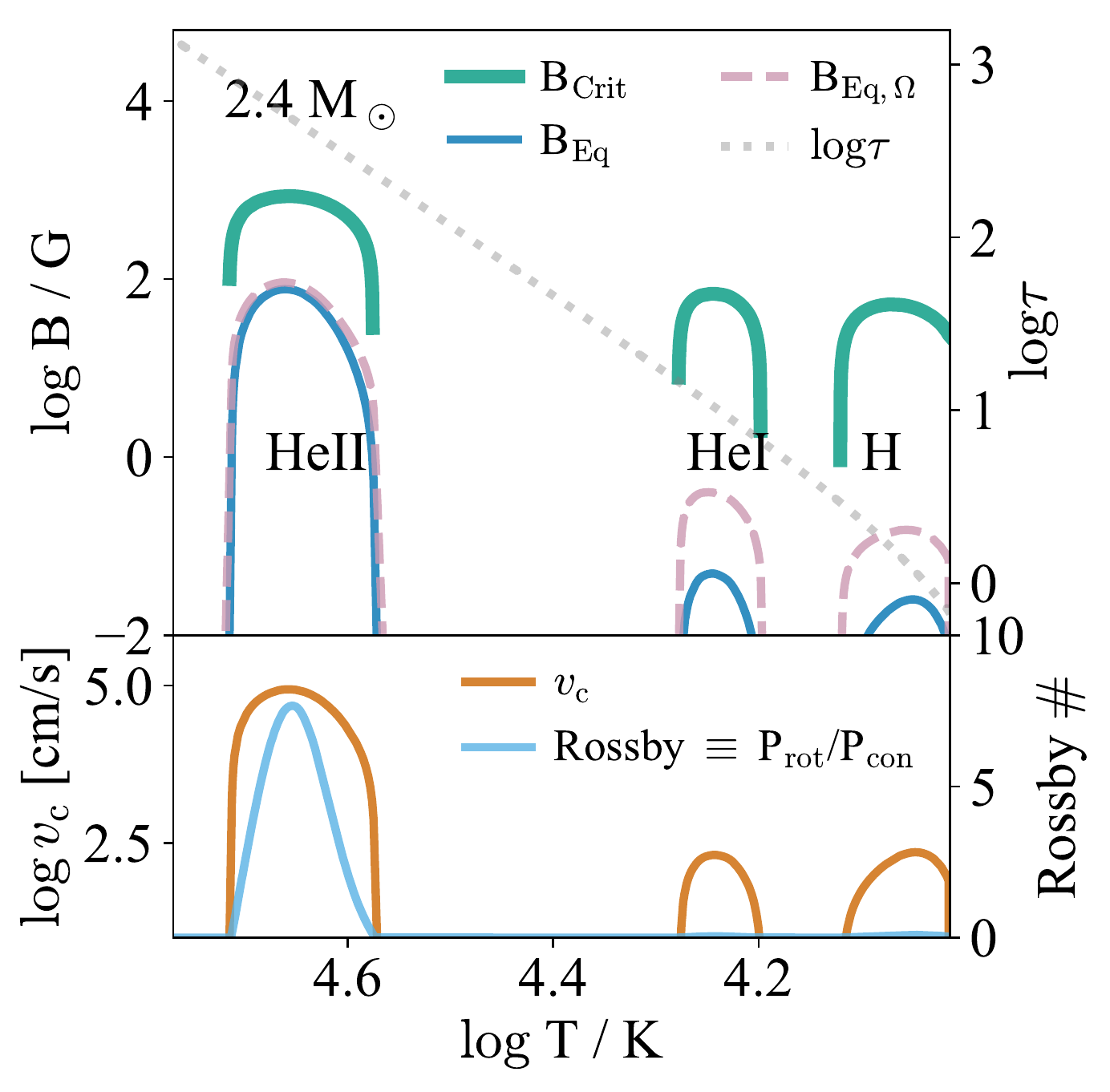}
\caption{(Upper) The critical magnetic field strength required to prevent convection (equation~\ref{eq:criterion}, green), the non-rotating dynamo saturation stregth (equation~\ref{eq:dynamo0}, blue), the same with a rotation rate of $150\,\mathrm{km\,s^{-1}}$ (equation~\ref{eq:dynamo}, pink), and the optical depth $\tau$ are shown as functions of $\log T/\mathrm{K}$ for a $2.4M_\odot$ stellar model. (Lower) The convection speed $v_{\rm c}$ and Rossby number ($P_{\rm rot} |N| / 2\pi$) are shown as functions of $\log T/\mathrm{K}$ on the same horizontal scale. The model was extracted at a fractional main sequence age of 0.79.}
\label{fig:Profiles_2.4}
\end{figure}

\begin{figure}
\centering
\includegraphics[width=0.47\textwidth]{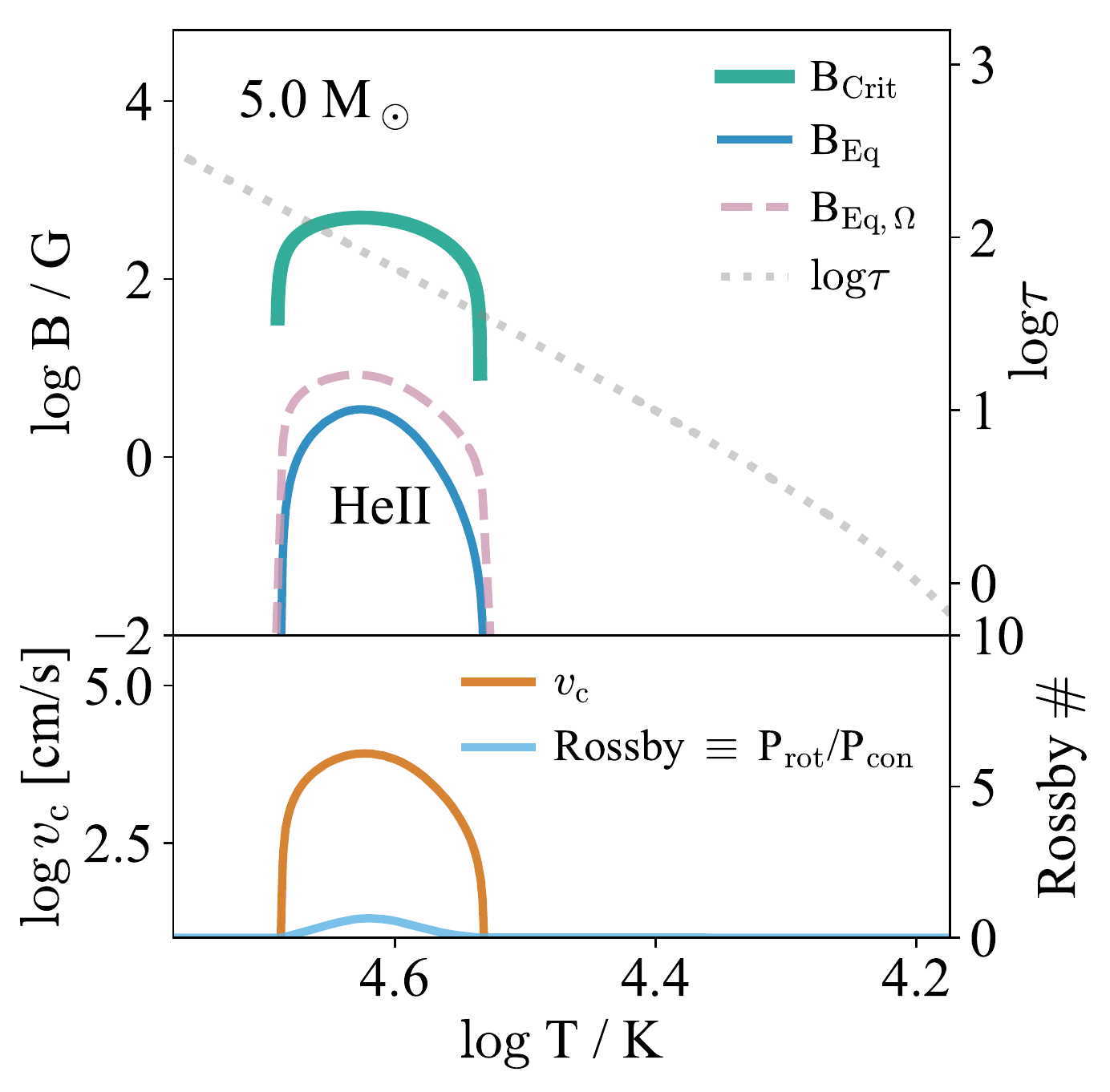}
\caption{Same as Fig.~\ref{fig:Profiles_2.4} but for a $5M_\odot$ stellar model at a fractional main sequence age of 0.77.}
\label{fig:Profiles_5.0}
\end{figure}

\begin{figure}
\centering
\includegraphics[width=0.47\textwidth]{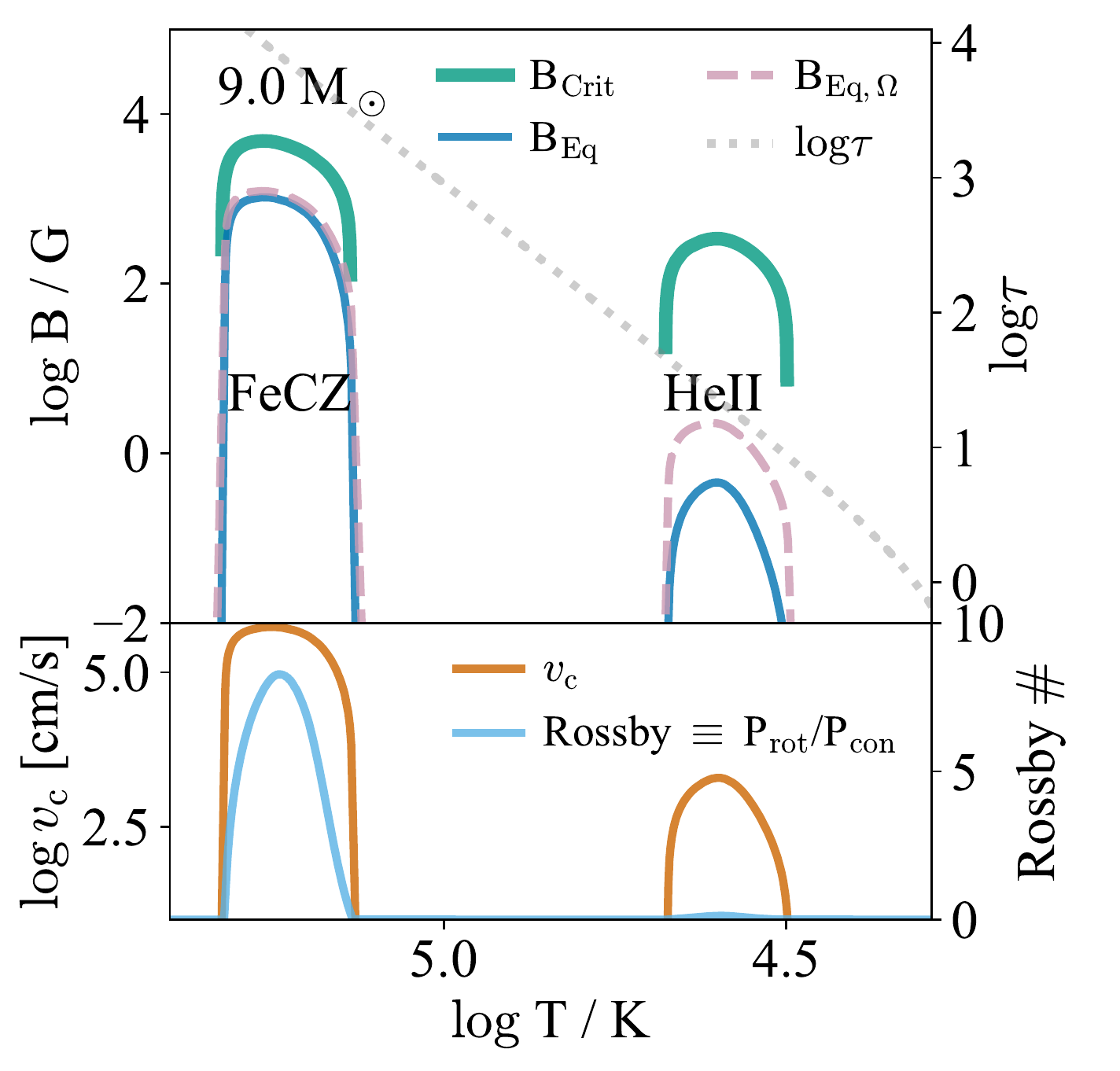}
\caption{Same as Fig.~\ref{fig:Profiles_2.4} but for a $9M_\odot$ stellar model at a fractional main sequence age of 0.95.}
\label{fig:Profiles_9.0}
\end{figure}

It is instructive to compare saturation field strength (equation~\ref{eq:dynamo0}) to the critical magnetic field (equation~\ref{eq:criterion}).
In the simple limit of an ideal gas with $v_{\rm A} \ll c_{\rm s}$, the ratio of these fields is
\begin{align}
	\frac{B_{\rm dynamo}}{B_{\rm crit}} = \frac{v_{\rm c}}{c_{\rm s}}\sqrt{\left(\nabla - \nabla_{\rm ad}\right)}.
\end{align}
Convection is inefficient in these stars, so $\sqrt{\nabla-\nabla_{\rm ad}}$ is of order unity and $v_{\rm c} \ll c_{\rm s}$, so the dynamo typically saturates at very sub-critical field strengths\footnote{This can change in the subsurface convection zones of very massive stars, where the large luminosity can drive turbulent velocities close to $c_{\rm s}$ \citep{2015ApJ...808L..31G,2015ApJ...813...74J,2018Natur.561..498J}}.
This may be seen in Fig.~\ref{fig:Profiles_2.4}, which shows the dynamo saturation field strength (equation~\ref{eq:dynamo0}) and the critical field strength $B_{\rm crit}$ as functions of temperature in the subsurface convection zones of a main-sequence $2.4M_\odot$ stellar model.
In all three subsurface convection zones the critical magnetic field is much stronger than the saturated dynamo field.
The same is true for a $5 M_\odot$ model (Fig.~\ref{fig:Profiles_5.0}) and a $9 M_\odot$ model (Fig.~\ref{fig:Profiles_9.0}), though in the latter case note that the FeCZ has a moderate Mach number ($\approx 0.07$) and so the two scales are only separated by a factor of a few. This more than compensates for the fact that the FeCZ has a lower superadiabaticity $\nabla - \nabla_{\rm ad}$.

In rotating stars the situation is more complex.
In numerical simulations the magnetic field has been shown to become super-equiparition once the Rossby number becomes small~\citep{doi:10.1111/j.1365-246X.2006.03009.x, 2017JFM...813..558A}.
Simulations by~\citet{2016ApJ...829...92A} suggest that in this limit
\begin{align}
	\frac{B^2}{\rho v_{\rm c}^2} \approx \frac{\Omega h}{v_{\rm c}},
\end{align}
where $\Omega$ is the angular velocity of the convection zone,
\begin{align}
	h \equiv \frac{P}{\rho g}
\end{align}
is the pressure scale height and $g$ is the acceleration owing to gravity.
Combining this scaling with the non-rotating limit in equation~\eqref{eq:dynamo0}, we find
\begin{align}
	B \approx \sqrt{4\pi \rho v_{\rm c}^2\left(1 + \frac{\Omega h}{v_{\rm c}}\right)}.
	\label{eq:dynamo}
\end{align}
In this rapidly-rotating limit the convection speed is also reduced by rotation~\citep{1982GApFD..21..113S,2020MNRAS.493.5233C}, so  if we evaluate equation~\eqref{eq:dynamo} with the non-rotating convection speed we should obtain an upper limit on the saturation strength of the convective dynamo.
An example of this upper limit is also shown in Figs.~\ref{fig:Profiles_2.4},~\ref{fig:Profiles_5.0}, and~\ref{fig:Profiles_9.0} (dashed) for models rotating at $150\,\mathrm{km\,s^{-1}}$. This is a typical value for the equatorial rotational velocity of OBA stars \citep{2008ApJ...683.1045H,2012A&A...537A.120Z,2014A&A...562A.135S}.
In the $2.4M_\odot$ HeII and $9 M_\odot$ Fe convection zones rotation is slow relative to convection, so the effect of rotation is minimal.
By contrast, the other four subsurface convection zones in these models are slow, so the effect of rotation is to increase $B_{\rm dynamo}$ by a factor of up to $3-10$.

Finally, the dynamo saturation field strength sets the amplitude of the magnetic field in the convection zone, but that is not a directly-observable quantity, so we must estimate the amplitude of the dynamo-driven field that emerges at the surface of the star.
The small-scale field produced by the dynamo field is expected to weaken by the time it reaches the surface.
\citet{2019ApJ...883..106C} argue that the minimum magnetic field at the surface should be
\begin{align}
	B_{\rm surface} \approx B_{\rm dynamo} \left(\frac{\rho_{\rm surface}}{\rho_{\rm CZ}}\right)^{2/3},
	\label{eq:B_surface}
\end{align}
where $\rho_{\rm CZ}$ is the density in the convection zone and $\rho_{\rm surface}$ is the density at the photosphere.

\begin{figure}
\centering
\includegraphics[width=0.47\textwidth]{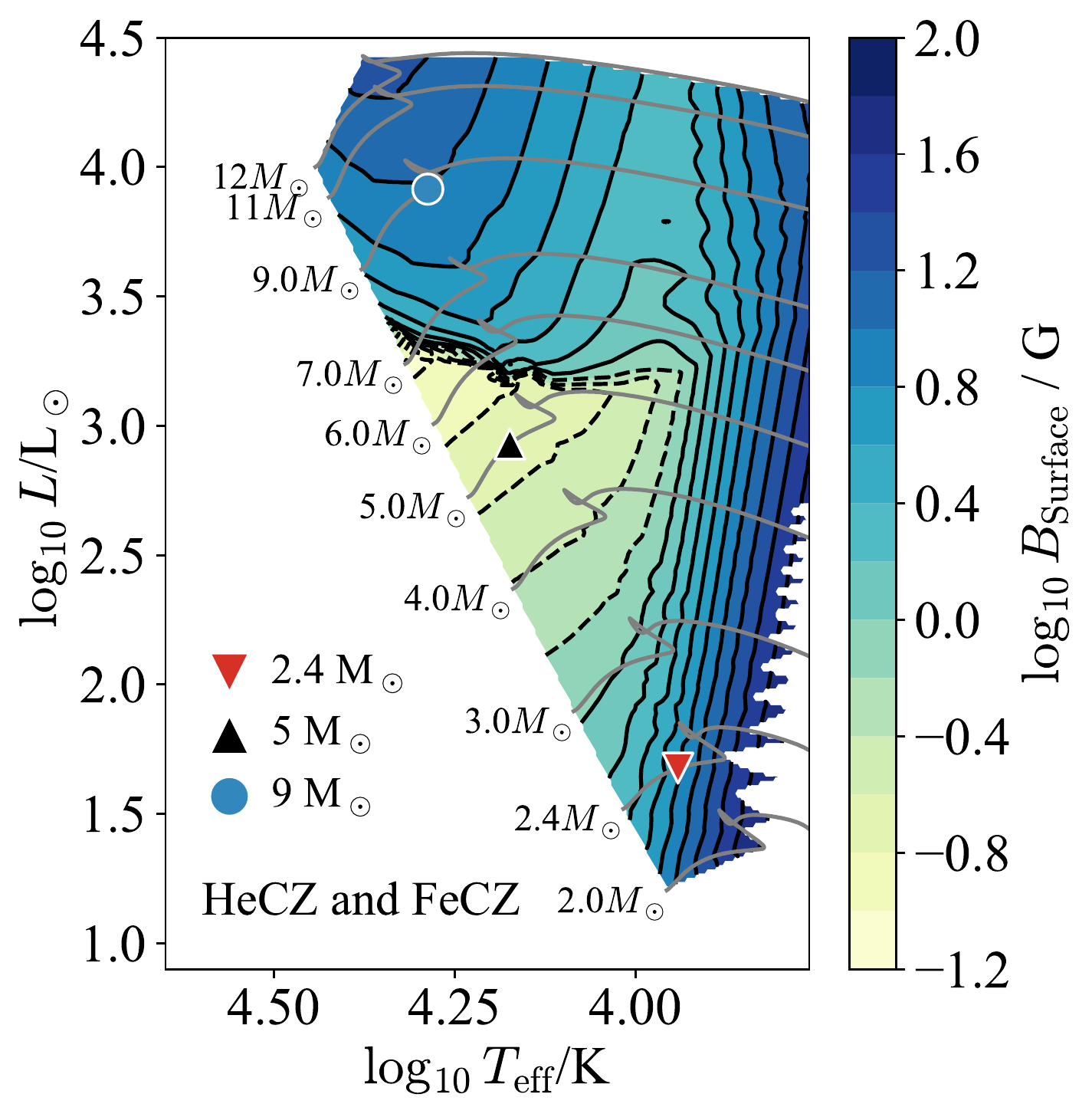}
\caption{The surface magnetic field $B_{\rm surface}$ from a dynamo in the dominant subsurface convection zones  is shown on a Hertzsprung-Russel diagram in terms of $\log T_{\rm eff}$ and $\log L$ for stellar models ranging from $2-12 M_\odot$. We only consider the HeCZ and FeCZ, and we plot values calculated using a radial average of the convective velocities. The surface values are calculated assuming a scaling $B\propto \rho^{2/3}$. The ridge at $\log L/L_\odot \approx 3.2$ is due to the appearance of the FeCZ. The interpolation artifact at low $\log T_{\rm eff}$ and low $\log L$ corresponds to the vigorous onset of H convection, rapidly moving deeper in the model and merging with the subsurface convection zones. We do not report values calculated for the H convection zone in this diagram.}
\label{fig:Bsurf}
\end{figure}

\begin{figure}
\centering
\includegraphics[width=0.47\textwidth]{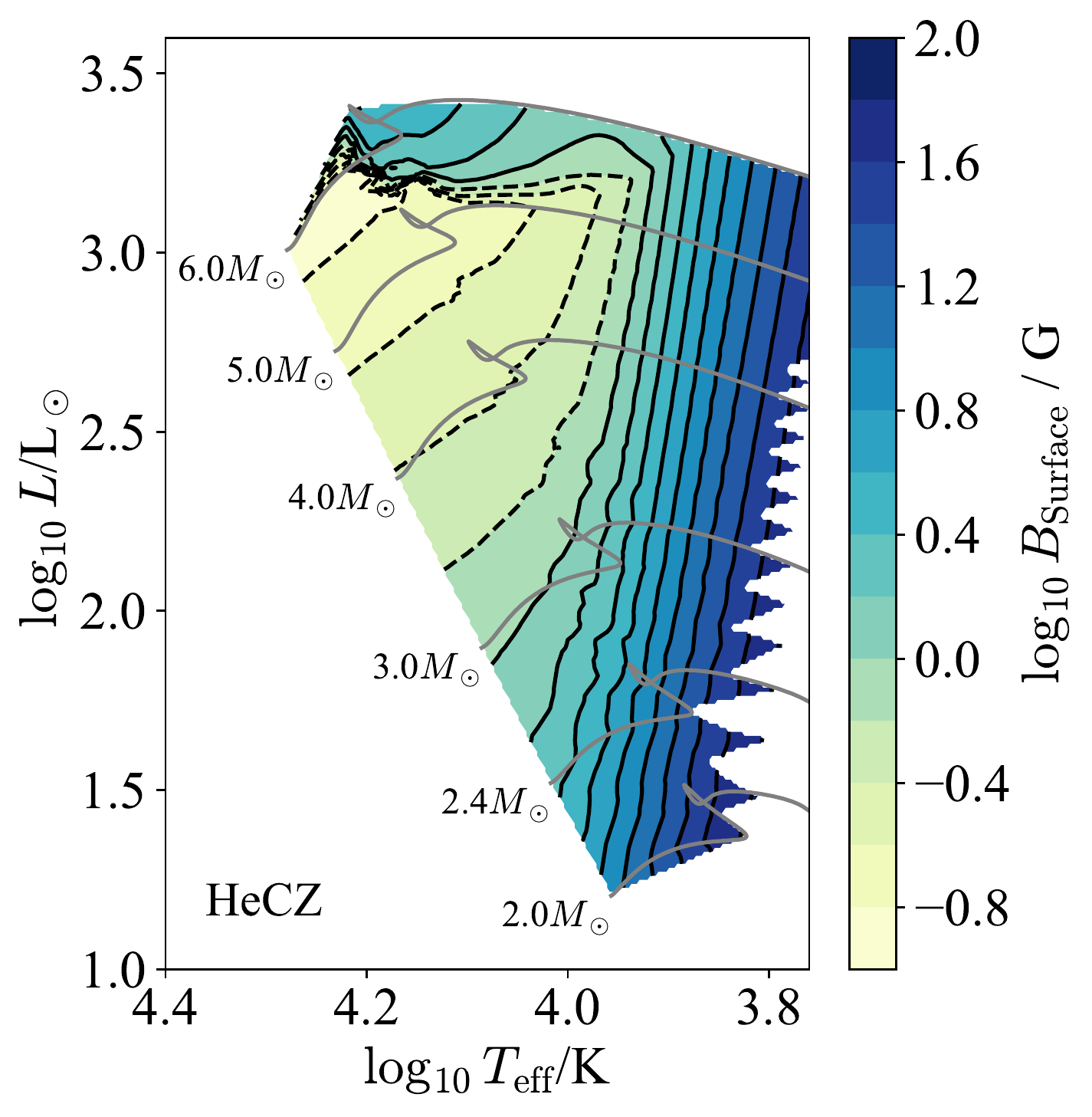}
\caption{The average surface magnetic field $B_{\rm surface}$ from a dynamo in the HeII convection zone is shown on a Hertzsprung-Russel diagram in terms of $\log T_{\rm eff}$ and $\log L$ for stellar models ranging from $2-6 M_\odot$. We plot values calculated using a radial average of the convective velocities. The surface values are calculated assuming a scaling $B\propto \rho^{2/3}$.}
\label{fig:Bsurf_HeCZ}
\end{figure}

We adopt this correction in Fig.~\ref{fig:Bsurf}, which shows the dynamo field adjusted using equation~\eqref{eq:B_surface} across the H-R diagram.
For $M > 7 M_\odot$ the surface field is dominated by the FeCZ, producing fields of $10-100\,\mathrm{G}$.
For lower masses the magnetic field is generated by the HeII convection zone (Fig.~\ref{fig:Bsurf_HeCZ}) and is considerably weaker, of order $0.1-1\,\mathrm{G}$.
Note that in both cases the surface field mostly increases with age, and this is particularly the case for stars which begin dominated by HeII convection and end dominated by the FeCZ.

Note that a major uncertainty in the absolute scale of the dynamo-driven magnetic field (equation~\ref{eq:dynamo}) arises because, for inefficient convection zones like the ones of interest, $v_{\rm c} \propto \alpha^3$, where $\alpha$ is the mixing length parameter~\citep{2019ApJ...883..106C}.
The uncertainty in this parameter is of order a factor of $2$, which translates into an order of magnitude uncertainty in the dynamo saturation field strength.

\section{Magnetic Evolution Model}
\label{sec:prediction}

We model the evolution of fossil fields through a simple physical argument.
When the fossil field strength $B_{\rm fossil} > B_{\rm crit}$, the fossil field is stable and convection is shut off.
When $B_{\rm fossil} < B_{\rm crit}$ convection is able to proceed.

\begin{figure}
\centering
\includegraphics[width=0.4\textwidth]{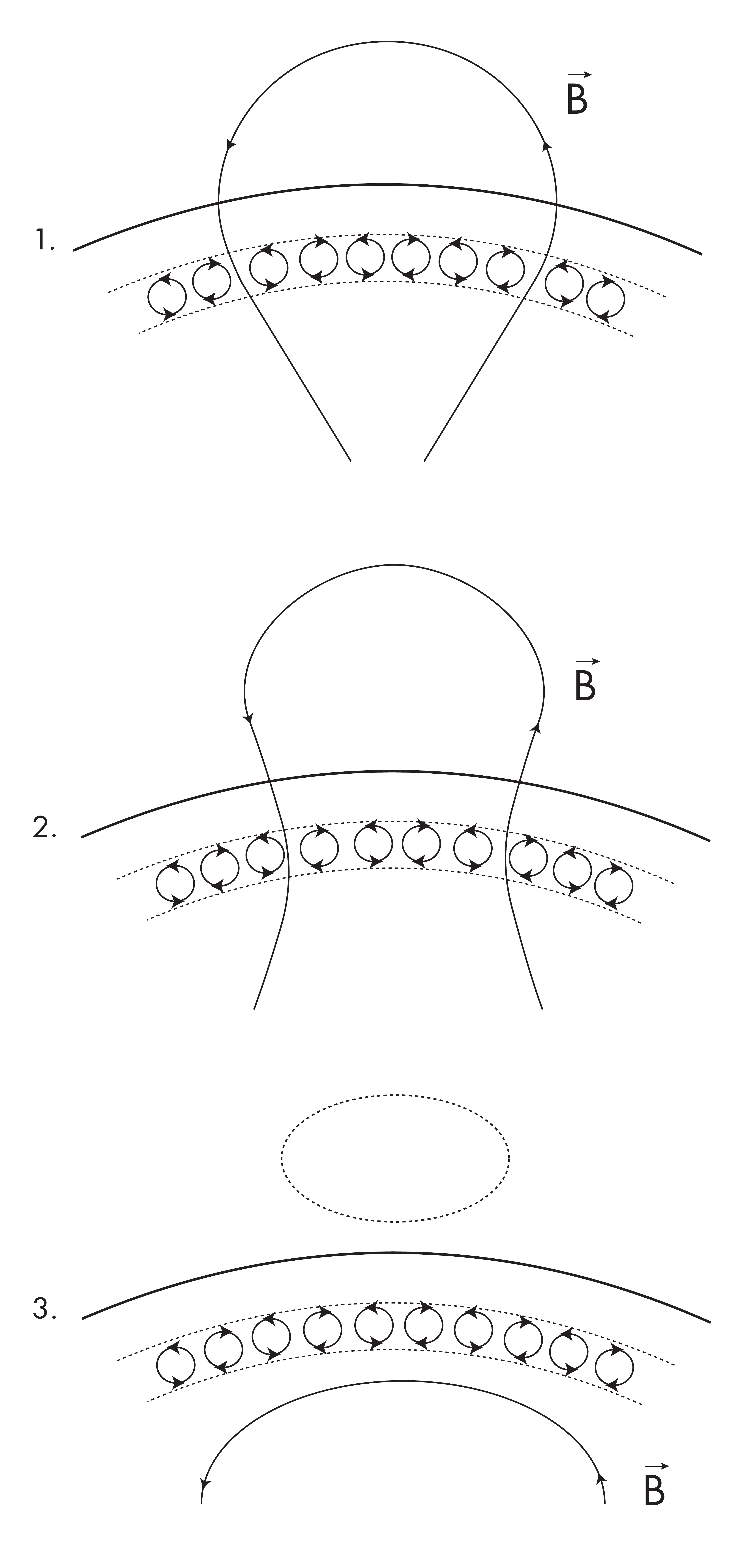}
\caption{A fossil field (upper, black) is shown twisted (middle) by convective motion, resulting in reconnection and ejection from the near-surface region (lower).}
\label{fig:schematic}
\end{figure}

Convection twists the fossil field on a turnover time-scale $\tau \approx |N|^{-1}$, where $N$ is the \brvs\ frequency, which is imaginary in a convection zone.
This destroys the large-scale structure of the field, pushing field energy towards ever-smaller scales until dissipation reduces the field strength to the dynamo saturation strength.
The net effect is to expel magnetic flux from the convection zone~\citep{Zeldovich,1989ZhETF..96..885A}.

The field in the surface radiative zone is unlikely to be left in a stable configuration and so reconnects on an Alfv{\'e}n time.
Even if it was in a stable configuration, in the shallow surface radiative zone of low-mass stars the field diffuses on a timescale $t_{\rm D} \approx H^2/\eta \approx 10^{4}\,\mathrm{yr}$, while for higher-mass stars it is expelled by winds on a time-scale of order $\Delta M / \dot{M} \approx 50\,\mathrm{yr}$~\citep{2011A&A...534A.140C}.

This process is shown schematically in Fig.~\ref{fig:schematic}.
Note that this need not alter the field in deeper regions significantly: it may simply be that the field within the convection zone is reduced, decoupling the field observed at the surface from that deeper in.
Moreover, for shallow convection zones this process likely only expels a fraction of the poloidal component of the field, since reconnecting the toroidal component requires bunching field lines close together on scales of order $h \ll r$ to re-route them through the convection zone. Since we assume that the magnetic field was initially stable, the outcome of this process is also a stable magnetic field configuration.
This is because reconnection events only occur in or above the convection zone, so that magnetic helicity is conserved in deeper regions.
In addition, the predominant loss of poloidal magnetic flux increases the ratio of toroidal magnetic energy to poloidal magnetic energy, increasing the configuration' stability~\citep{2009MNRAS.397..763B}.
The toroidal component is not directly observable though, so in stars which have undergone this process it is possible that a toroidal component remains in the convection zone.

\begin{figure}
\centering
\includegraphics[width=0.46\textwidth]{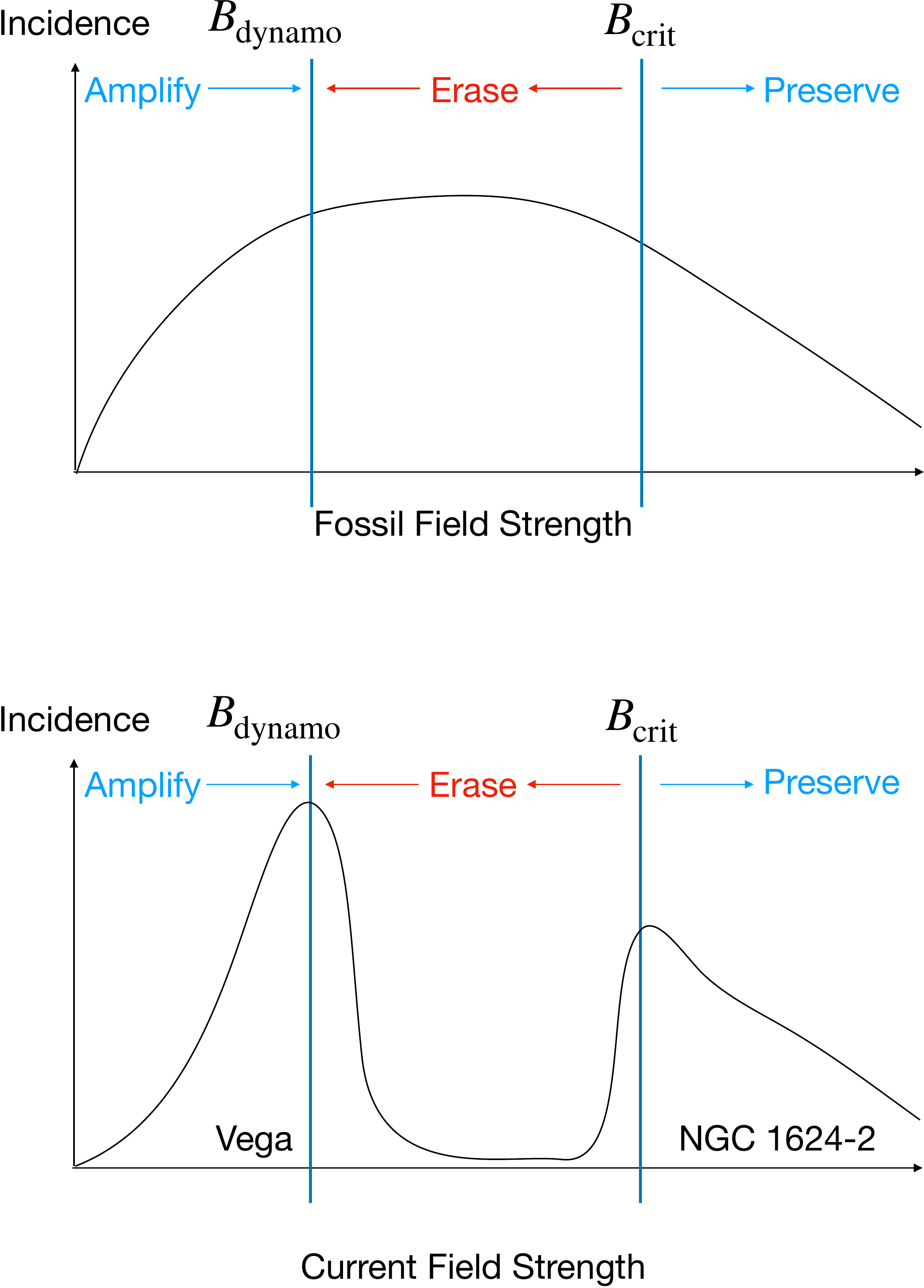}
\caption{The evolution of the distribution of stellar magnetic field strength is shown schematically. Initially there is a broad distribution. Fields stronger than the critical field strength are preserved. Stars with weaker magnetic fields see theirs either amplified ($B_{\rm fossil} < B_{\rm dynamo}$) or erased ($B_{\rm fossil} > B_{\rm dynamo}$), resulting in a pileup at $B_{\rm dynamo}$. Vega and NGC~1624-2 are provided on the lower panel as examples of stars with dynamo-driven and fossil magnetic fields. The amplification or erasure takes place on a time-scale of order the convective turnover time. However, an unstable magnetic field may remain above a convection zone for an ohmic diffusion time (of order $10^{4}\,\mathrm{yr}$) or for more massive stars the time it takes for stellar winds to expel the near-surface material (of order $50\,\mathrm{yr}$), and so the observable surface field should only decay on the shorter of these time-scales.}
\label{fig:time_schematic}
\end{figure}

Because $\tau$ is much less than the main-sequence lifetime of the star, the result is that a sub-critical fossil field is rapidly erased and replaced with a less structured field on the order of the dynamo saturation stregth (equation~\ref{eq:dynamo}).
Thus we predict a bimodal distribution of field strengths, with each mode pointing to a distinct magnetic field origin.
This is shown schematically in Fig.~\ref{fig:time_schematic}.

Our predictions based on this analysis are that:
\begin{enumerate}
\item  Early-type stars should either have a weak magnetic field owing to a dynamo, or a strong fossil field above the largest $B_{\rm crit}$ encountered at any point in its evolutionary history.
\item Because $B_{\rm crit}$ generally decreases with increasing mass for $M < 7 M_\odot$, in this mass range we expect to see a higher fraction of stars strongly magnetized at higher masses.
\item Because $B_{\rm crit}$ initially increases with stellar age, we expect the fraction of stars with strong magnetic fields to decline with age. This is in addition to the effect where flux conservation causes a frozen-in magnetic field to weaken as the star ages and expands.
\item Following~\citet{2019MNRAS.487.3904M}, if macroturbulence is caused by subsurface convection then stars with $B > B_{\rm crit}$ ought to have little or no macroturbulence.
\item As shown in Appendix~\ref{appen:grids}, $B_{\rm crit}$ increases with increasing metallicity, so we expect to see fewer stars with strong fossil fields at higher metallicity. Similarly, $B_{\rm dynamo}$ increases with increasing metallicity, so we expect to see stronger dynamo-driven fields at higher metallicity.
\end{enumerate}

Note that while this model is physically distinct from the magnetic instability model of~\citet{2007A&A...475.1053A}, the two are similar in that both invoke a critical magnetic field above which fossil fields are stable against the instability of interest.
In our case, however, the instability is convective and so we expect $B_{\rm crit}$ to be principally a function of the thermal structure of the star, whereas in their model this depends on differential rotation and so the scale potentially varies with rotation rate and rotational history even among stars of the same mass and age.
We thus expect the two models to predict different trends of magnetization with mass, age, and rotation, though we leave a detailed analysis of the differences for the future.

Finally \citet{2013MNRAS.428.2789B} discussed the possibility that some of the observed magnetic fields in rotating early-type stars might be
failed fossils. These fields are not in a stable configuration (like in Ap and Bp stars), but they still evolve relatively slowly thanks to the balance of Coriolis and Lorentz force in the momentum equation. Failed fossils are also expected to be erased by subsurface convection for amplitudes $B < B_{\rm crit}$, so  our conclusions about a dichotomy in the origin of magnetic fields and the resulting magnetic desert are independent on the specific fossil field scenario adopted.  Below subsurface convection zones, subcritical failed fossil fields can still exist and evolve slowly, and have an impact on stellar interior properties.

\section{Observations}
\label{sec:obs}

We now compare our predictions to observations.

\subsection{Bimodal Field Strengths}

First, we predict a bimodal distribution of magentic field strengths.
This is generally what is seen~\citep{2014IAUS..302..338L}.

\citet{2007A&A...475.1053A} detected magnetic fields in $28$ Ap stars with $\log T_{\rm eff}/\mathrm{K}$ ranging from $3.9-4.1$ and $\log L/L_\odot$ from $1.1-2.7$.
For these stars we predict a critical field strength of order $700\,\mathrm{G}$.
Given that the Ap phenomenon is believed to be a result of strong magnetic fields altering the chemical mixing of the star~\citep{1993ASPC...44..458B}, we expect most of this sample to exhibit fossil fields with strengths of order $B_{\rm crit}$ or larger.
Indeed all but one of the other stars in their sample have best-fit field strengths above $200\,\mathrm{G}$, all but two are consistent with $B > 700\,\mathrm{G}$, and half of the sample have best-fit field strengths greater than $700\,\mathrm{G}$.

At the other end of the spectrum,~\citet{2010A&A...523A..40A} placed upper limits of order $5\,\mathrm{G}$ on the typical magnetic fields of A stars, while spot mesaurements~\citep{2017MNRAS.467.1830B,2019MNRAS.490.2112B,2020MNRAS.492.3143T} indicate that most of these stars do have some weak magnetic field.
Likewise~\citet{2010A&A...523A..41P} detected a magnetic field strength of order $1\,\mathrm{G}$ in Vega and~\citet{2011A&A...532L..13P} detected a field strength of $0.2\,\mathrm{G}$ in Sirius A.
These weak fields are consistent with what we expect.
For A/B stars the dynamo saturation strength is of order $10-100\,\mathrm{G}$ and the emergent surface flux is of order $1-10\,\mathrm{G}$~\citep{2019ApJ...883..106C}.
Moreover, except for rapid rotators we expect the convective dynamo to primarily generate a small-scale magnetic field with almost no dipole component, so it is consistent that of the positive detections the more rapidly rotating star (Vega) exhibits a stronger large-scale field.

At the high-mass end,~\citet{2015A&A...574A..20F} report the detection of a $60-230\,\mathrm{G}$ field in $\beta$CMa, and a lower limit of $13\,\mathrm{G}$ for $\epsilon$CMa.
Despite their claim that this is inconsistent with a magnetic desert model, these observations are consistent with what we expect if the fields are dynamo-generated, though it is perhaps somewhat surprising that so much power lies in the dipole mode given their moderate rotation ($v\sin i \approx 20\,\mathrm{km\,s^{-1}}$).
$\beta$CMa and $\epsilon$CMa are extremely luminous, massive stars with $M \approx 12 M_\odot$, $\log L/L_\odot \approx 4.4$, and $\log T_{\rm eff}/\mathrm{K} \approx 4.4$.
If their magnetic fields are generated by subsurface convective dynamos, the expected surface field strength is of order $30\,\mathrm{G}$, which is consistent with observations of $\epsilon$CMa and not far from what is observed in $\beta$CMa, especially given the uncertainties in deriving equipartition magnetic fields from mixing length theory~\citep{2019ApJ...883..106C}.

We feel compelled to mention that despite this general agreement with our model, it is possible that the magnetic desert results simply from observational incompleteness~\citep{2010AstL...36..370K}.
This makes our further predictions especially salient, as they are less susceptible to this difficulty.

\subsection{Mass Distribution}

Because $B_{\rm crit}$ generally decreases with increasing mass for $M < 7 M_\odot$, also we expect to see a higher fraction of stars strongly magnetized at higher masses.
\citet{2019MNRAS.483.3127S} report a volume-limited sample of 52 chemically peculiar A and B stars, and find that the fraction of detectable magnetic fields rises from $10^{-2}$ at $M=2 M_\odot$ to over $10$~per-cent at $3-4 M_\odot$.
A majority of the increase in magnetic fraction lies between $2.2 M_\odot$ and $3 M_\odot$.
If our model for the origins of these magnetic fields is correct, and if the typical fossil field strength of chemically peculiar stars is independent of stellar mass, this suggests they lie mostly between $B_{\rm crit}(3 M_\odot) \approx 700\,\mathrm{G}$ and $B_{\rm crit}(2.2 M_\odot) \approx 900\,\mathrm{G}$.

At higher masses,~\citet{2010AstL...36..370K} report that the fraction of O stars with measured magnetic fields is $1/3$ of that of B stars.
This is consistent with $B_{\rm crit}$ rising by $10-30$ fold in the mass range $5 M_\odot < M < 7 M_\odot$.
If the initial fossil field strength in O stars is similar to that in B stars, this increase means that convection zones erase the fossil fileds in a much greater fraction of O stars than B stars, resulting in weaker fields and more difficult detections.

\subsection{Age Distribution}

\begin{figure}
\centering
\includegraphics[width=0.47\textwidth]{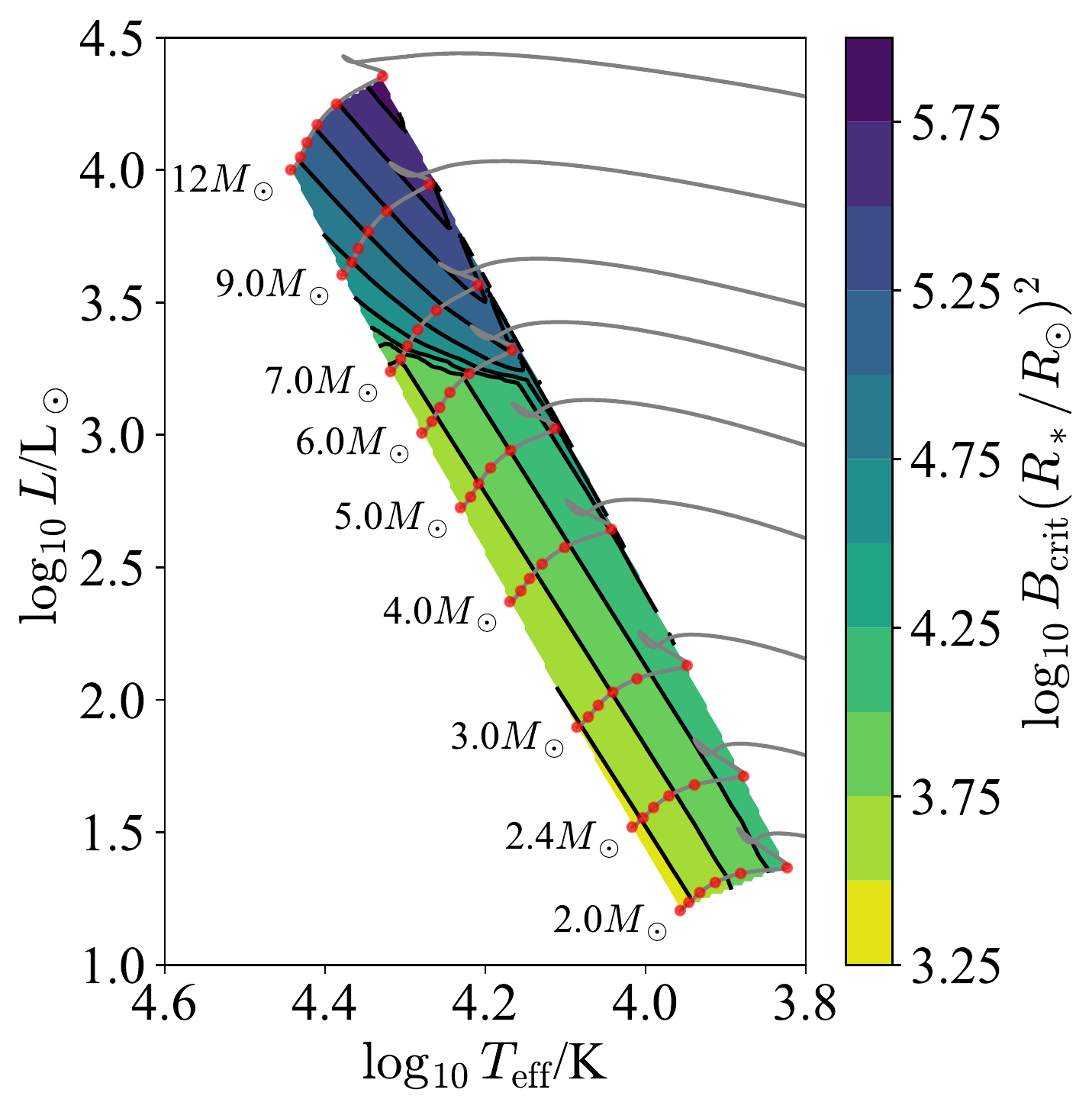}
\caption{The critical magnetic flux $B_{\rm crit} (R/R_\odot)^2$ is shown on a Hertzsprung-Russel diagram in terms of $\log T_{\rm eff}$ and $\log L$ for stellar models ranging from $2-12 M_\odot$ with Z=0.02. We do not report values calculated for the H convection zone in this diagram. Red dots show the location of 20\% increase in fractional age, from the zero age main sequence to hydrogen exhaustion.}
\label{fig:critical_flux}
\end{figure}

Because $B_{\rm crit}$ initially increases with stellar age, we expect the fraction of stars with strong magnetic fields to decline with age.
Moreover the radius of the star increases with age, so if magnetic flux is conserved the fossil magnetic field should decrease in strength as the star ages, pushing ever-more stars below the critical field strength and erasing their fossil fields.
The combined effect of $B_{\rm crit}$ increasing and $R$ increasing with time results in $B/B_{\rm crit}$ falling by a factor of $3-10$ over the main-sequence evolution of these stars (Fig.~\ref{fig:critical_flux}).

This prediction is consistent with observations of O and B stars~\citep{2016A&A...592A..84F}, which suggest that the magnetic fraction declines starting around $0.4$ of the main-sequence lifetime, where it is of order $15$~per-cent, to nearly $0$ by the end of the main sequence.
Along similar lines,~\citet{2010AstL...36..370K} found that the population-averaged magnetic field of O and B stars declines with age like $e^{-2 t / \tau}$, where $\tau$ is the main-sequence stellar lifetime.
Some of this decline can be attributed to flux conservation as the stars expand, which could produce a factor of a few, though it is likely that this does not explain the full decline.

That the decline in field strength happens on the evolutionary time-scale of the star despite the wide mass range considered suggests that it is a matter of stellar structure as our model predicts, rather than being set by the magnetic diffusion time.
\citet{2018CoSka..48..223M} find a similar decrease in the mean magnetic field strength with age.
Their figure~9 shows that while there is a slight decrease in the strongest O/B stellar magnetic fields with age, the bulk of the reduction in mean magnetic field arises due to the appearance of O stars with weak ($30-300\,\mathrm{G}$) magnetic fields around $t/\tau \approx 0.5-1$.
This is consistent with the dynamo fields of less massive stars being too weak to detect, and with a sub-population of O stars developing subsurface convection zones and replacing their fossil magnetic fields with much weaker equipartition ones.

Other lines of evidence point in the same direction.
\citet{2007A&A...466..269B} report that Bp stars are on average much younger and more strongly magnetized than SPB stars, suggesting either that magnetic fields interfere with the pulsations of SPB stars or that many Bp stars lose their magnetic fields en route to becoming SPB stars.
The former possibility is inconsistent with observations of the SPB star o~Lup with $B \approx 5250\,\mathrm{G}$~\citep{2018arXiv180805503B}, so we think it likely that the fossil fields of a substantial fraction of B stars are erased by the time they reach the SPB phase.

\subsection{Macroturbulence}

If macroturbulence is a result of subsurface convective motions, we expect stars with $B > B_{\rm crit}$ to show little or no evidence of macroturbulence\footnote{Similarly, since microturbulence is believed to be caused by subsurface convection \citep{2009A&A...499..279C}, it should be absent or negligible in stars with $B > B_{\rm crit}$}.
In a similar analysis for more massive stars, first~\citet{10.1093/mnras/stt921} and later~\citet{2019MNRAS.487.3904M} suggest this as the reason that NGC~1624--2 lacks macroturbulence, and with a measured field strength of $16-20\,\mathrm{kG}$ it is a good candidate for this effect.

HD~215441 (Babcock's star) provides a similar example, with $T_{\rm eff} \approx 14,500\,\mathrm{K}$ and $B \approx 67\,\mathrm{kG}$.
\citet{1989ApJ...344..876L} report that the spectrum of HD~215441 requires macroturbulence $\xi > 3\,\mathrm{km\,s^{-1}}$ and is consistent with zero macroturbulence.

The other candidate of which we are aware is HD~54879.
This star has $T_{\rm eff} \approx 33,000\,\mathrm{K}$ and is strongly magnetized, with mean longitudinal field $B = -583\pm 9\,\mathrm{G}$.
Like NGC~1624--2, HD~54879 also exhibits weak macroturbulence: $\xi = 4 \pm 1\,\mathrm{km\,s^{-1}}$ from HARPS and $5\pm 3\,\mathrm{km\,s^{-1}}$ from FORS~2~\citep{2015A&A...581A..81C}, compared with more typical values of $20-60\,\mathrm{km\,s^{-1}}$~\citep{10.1093/mnras/stt921,2017A&A...597A..22S}.

\section{Discussion}
\label{sec:discussion}

We predict that strong magnetic fields in early-type stars are fossil fields, and that weak magnetic fields in these stars emerge from dynamo action in subsurface convection zones.
In our model these convection zones serve to erase any near-surface evidence of fossil fields by twisting them down to small length-scales where they may be dissipated.
An important task for future work is to test this basic physical picture in numerical simulations.

If correct, this prediction means that the observed bimodal distribution of magnetic fields is really an indication of two populations: one in which the magnetic field was strong enough to prevent near-surface convection, and one in which it was not.
These populations ought to exhibit very different surface magnetic field evolution, and may appear qualitatively different in terms of related near-surface phenomena like macroturbulence and microturbulence.

Our scenario may be distinguished from that of~\citet{2007A&A...475.1053A} by noting that their critical magnetic field strength is dependent on the stellar rotation rate while ours is not.
As a result their model predicts that the population of strongly-magnetized stars shifts to weaker field strengths as the rotation period increases while our model predicts no dependence on stellar rotation rate.
This comparison is potentially complicated in practice because magnetic breaking should reduce stellar rotation rates, but we are hopeful that these effects may be disentangled by comparing population synthesis models produced with our scenario and that of~\citet{2007A&A...475.1053A}.

Recently low-frequency variability has been detected in massive stars~\citep{2019NatAs...3..760B}.
There are competing explanations for this phenomenon, including internal gravity waves emitted by core convection~\citep{2019NatAs...3..760B} and motions excited by subsurface convection~\citep{2019ApJ...886L..15L}.
If the same phenomenon is detected in any star with a magnetic field $B > B_{\rm crit}$ that would be strong evidence against the hypothesis that the variability originates in a subsurface convection zone.
Conversely, an absence of low-frequency variability in strongly magnetized massive stars could point to an origin in subsurface convection zones.

Our model also has consequences for angular momentum transport.
While we posit that subsurface convection erases any subcritical fossil field at the surface, it likely leaves any magnetic field in the deeper interior largely unaltered.
Given the large fraction of early-B/late-O type stars with magnetic fields~\citep{2007pms..conf...89P,2015A&A...582A..45F,2017MNRAS.465.2432G}, it seems plausible that a large fraction of early-type stars without significant surface magnetism could still be strongly magnetized below the subsurface convection layers.

A strong fossil magnetic field hidden in the interior of a star could be subcritical yet still play an important role in angular momentum transport, potentially enforcing nearly rigid rotation via magnetic tension.
For instance a field strength of $10\,\mathrm{G}$ in a medium of typical stellar density $\rho \approx 0.1\,\mathrm{g\,cm^{-3}}$ suffices to generate a specific torque of order $10^3\,\mathrm{erg\,g^{-1}}$, which yields an angular acceleration over $r \sim 3\times 10^{11}\,\mathrm{cm}$ of $10^{-20}\,\mathrm{s^{-2}} \approx \Omega_{\odot}/\mathrm{Myr}$, where $\Omega_{\odot}$ is the mean angular velocity of the Sun.
So over the main-sequence lifetime of a massive star even such a weak and highly-subcritical field is enough to redistribute the entire angular momentum of the star many times over.

Conversely, differential rotation can amplify magnetic fields via the Spruit-Tayler dynamo~\citep{2002A&A...381..923S,2019MNRAS.485.3661F}.
With significant (order unity) differential rotation this can generate magnetic fields comparable to $B_{\rm crit}$ and so could provide another source of supercritical fossil fields, shutting off convection if the dynamo is active before subsurface convection layers form.

Finally, in stars with multiple subsurface convection zones, it is possible that the magnetic field from the FeCZ is strong enough to shut off the weaker overlying HeCZ.
Because these dynamo-generated magnetic fields have significant power at small scales, this likely does not happen everywhere in the HeCZ at the same time and so manifests with patches of active Helium-driven convection and patches of quiescence.

\acknowledgments

%%%%%%%%%%% jump to Snippets/acknowledgements
%%%---------- open: Snippets/acknowledgements
The Flatiron Institute is supported by the Simons Foundation.%%%---------- close: Snippets/acknowledgements
We are grateful to Daniel Lecoanet and Keaton Burns for helpful discussions on subsurface convection. We also thank Yuri Levin and Lars Bildsten for helpful comments on magnetic field configurations, as well as Pablo Marchant and Fabian Schneider for useful comments and some early conversations that helped spark some of the ideas in this work. MC thanks Maria Di Paolo for helping with producing the schematic illustration in section~\ref{sec:prediction}.
This research was supported in part by the National Science Foundation under Grant No. NSF PHY-1748958.

\appendix

\section{MESA Microphysics}
\label{appen:mesa}

All calculations were done with MESA version 11701.
The MESA EOS is a blend of the OPAL \citet{Rogers2002}, SCVH
\citet{Saumon1995}, PTEH \citet{Pols1995}, HELM
\citet{Timmes2000}, and PC \citet{Potekhin2010} EOSes.

Radiative opacities are primarily from OPAL \citep{Iglesias1993,
Iglesias1996}, with low-temperature data from \citet{Ferguson2005}
and the high-temperature, Compton-scattering dominated regime by
\citet{Buchler1976}.  Electron conduction opacities are from
\citet{Cassisi2007}.

Nuclear reaction rates are a combination of rates from
NACRE \citep{Angulo1999}, JINA REACLIB \citep{Cyburt2010}, plus
additional tabulated weak reaction rates \citet{Fuller1985, Oda1994,
Langanke2000}.  (For MESA versions before 11701): Screening is
included via the prescriptions of \citet{Salpeter1954, Dewitt1973,
Alastuey1978, Itoh1979}. (For MESA versions 11701 or later): Screening
is included via the prescription of \citet{Chugunov2007}.  Thermal
neutrino loss rates are from \citet{Itoh1996}.

\section{Effects of Suppressing Convection}
\label{appen:comparison}
\begin{figure}
\centering
\includegraphics[width=0.47\textwidth]{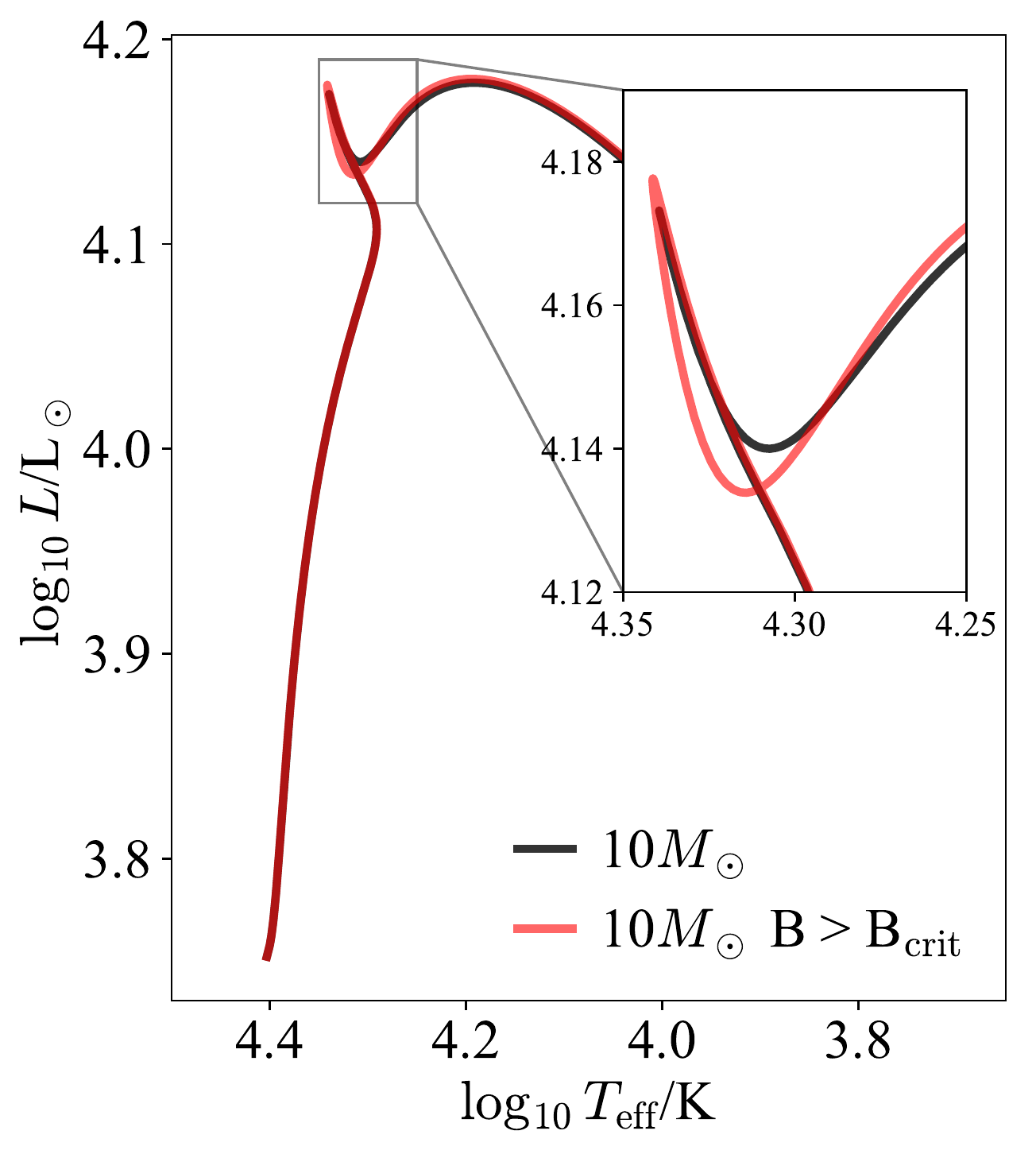}
\caption{Impact of suppressing subsurface convection on the evolution of a 10$M_\odot$ model. To mimic the presence of a magnetic field with $B>B_{\rm crit}$, for $T < 500,000 $ K we force the flux to be exclusively radiative. The effect is negligible during the main sequence.}
\label{fig:10Suppressed}
\end{figure}

Here we look at the possible effects of suppressing subsurface convection zones on the effective temperature and luminosity of stars.
The suppression of subsurface convection is achieved by forcing the flux to be purely radiative below a temperature of 500,000 K. This is to mimic the effect of a magnetic field with amplitude  $B > B_{\rm crit}$. Fig.~\ref{fig:10Suppressed} shows that the effect of shutting off the subsurface convective regions in a 10$M_\odot$ model is pretty much negligible. This is expected, since the flux carried by convection in these regions is very small \citep[Usually 1\% or less, see e.g. Fig.6 in ][]{2019ApJ...883..106C}.

\section{Grids at Different Metallicities}
\label{appen:grids}
Here we present results for the critical and surface magnetic field at metallicities of Z=0.014, Z=0.006, and Z=0.002,
representing early-type stars in the Galaxy (MW), Large Magellanic Cloud (LMC), and Small Magellanic Cloud (SMC) respectively  \citep[e.g.][]{2013MNRAS.433.1114Y}. The initial Helium content of the grids is
Y=0.2659 (MW), Y=0.2559 (LMC), and Y=0.2508 (SMC). The metallicity is initialized scaling the standard solar composition of \citet{1998SSRv...85..161G}.

\begin{figure}
    \centering
    \begin{minipage}{0.48\textwidth}
        \centering
        \includegraphics[width=1\textwidth]{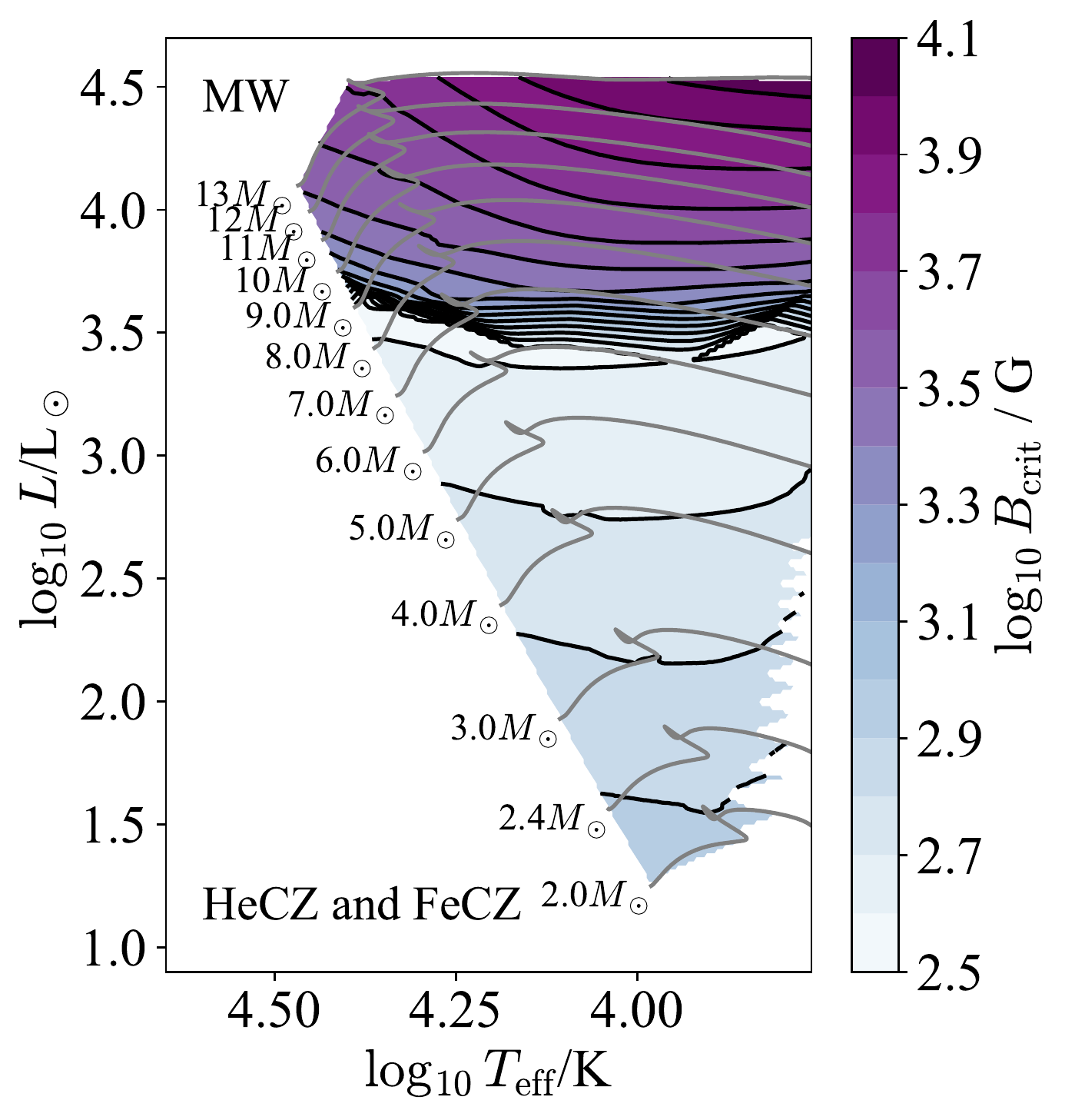} % first figure itself

    \end{minipage}\hfill
    \begin{minipage}{0.5\textwidth}
        \centering
        \includegraphics[width=1\textwidth]{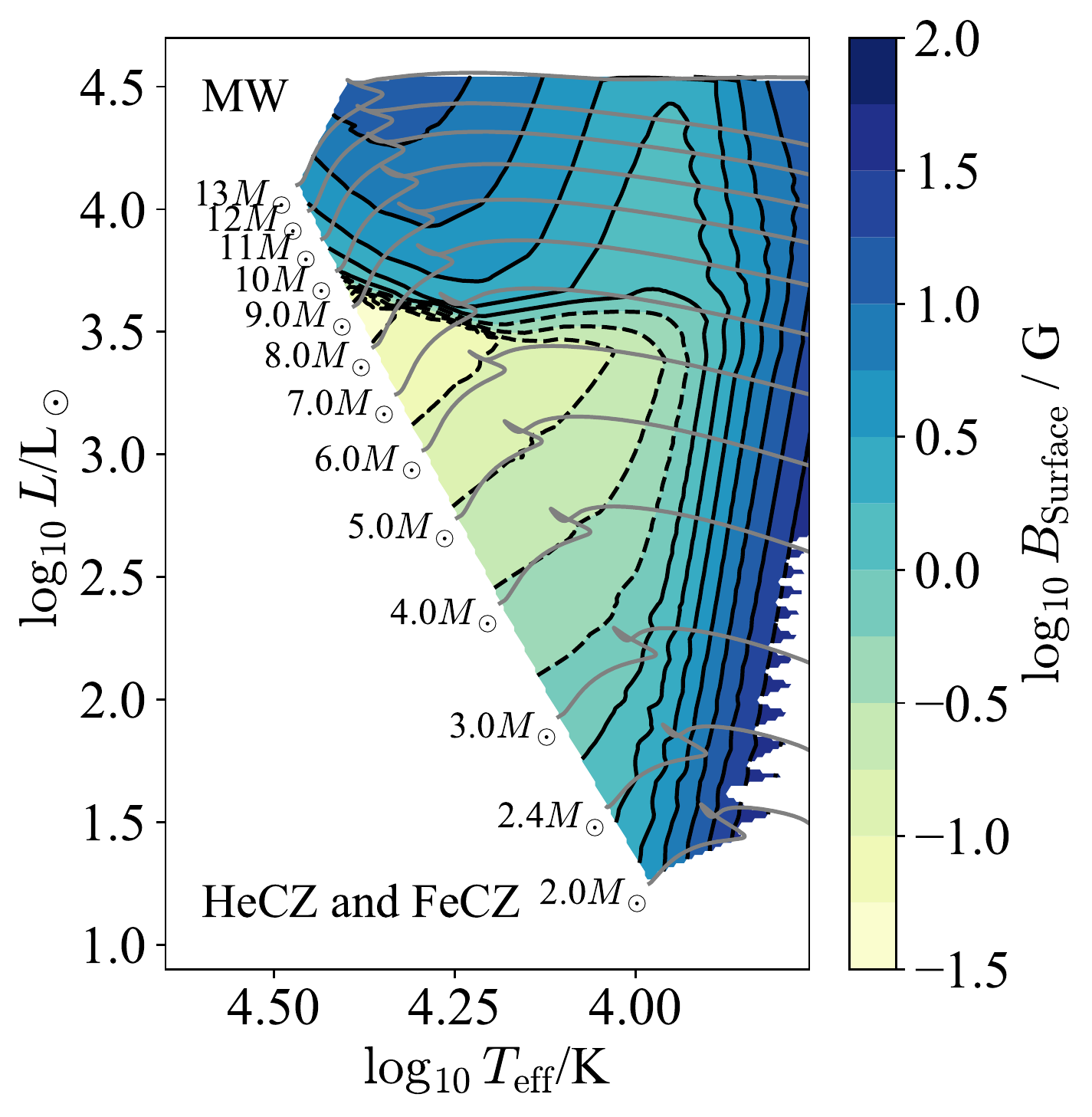} % second figure itself
    \end{minipage} \caption{Left: The critical magnetic field $B_{\rm crit}$ given by equation~\eqref{eq:criterion} is shown on a Hertzsprung-Russel diagram in terms of $\log T_{\rm eff}$ and $\log L$ for stellar models ranging from $2-13 M_\odot$ with Z = 0.014. The ridge at $\log L/L_\odot \approx 3.5$ is due to the appearance of the FeCZ. The interpolation artifact at low $\log T_{\rm eff}$ and low $\log L$ corresponds to the vigorous onset of H convection, rapidly moving deeper in the model and merging with the subsurface convection zones. We do not report values calculated for the H convection zone in this diagram. Right: The surface magnetic field $B_{\rm surface}$ from a dynamo in the dominant subsurface convection zones  is shown on a Hertzsprung-Russel diagram in terms of $\log T_{\rm eff}$ and $\log L$ for stellar models ranging from $2-13 M_\odot$ with Z = 0.014. We only consider the HeCZ and FeCZ, and we plot values calculated using a radial average of the convective velocities. The surface values are calculated assuming a scaling $B\propto \rho^{2/3}$. The ridge at $\log L/L_\odot \approx 3.2$ is due to the appearance of the FeCZ. The interpolation artifact at low $\log T_{\rm eff}$ and low $\log L$ corresponds to the vigorous onset of H convection, rapidly moving deeper in the model and merging with the subsurface convection zones. We do not report values calculated for the H convection zone in this diagram.}\label{fig:MW_Bcrit}
\end{figure}

\begin{figure}
    \centering
    \begin{minipage}{0.48\textwidth}
        \centering
        \includegraphics[width=1\textwidth]{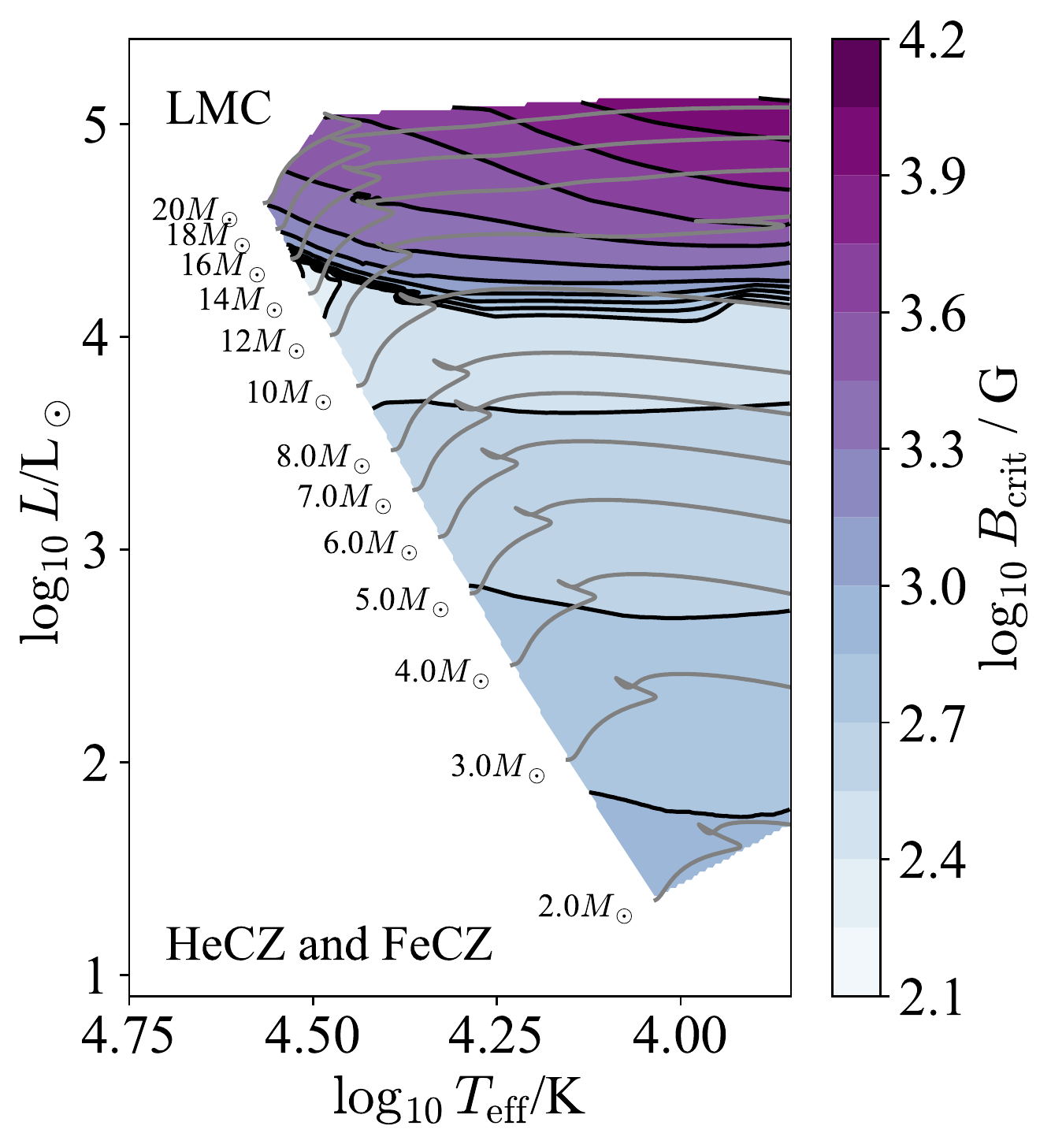} % first figure itself
    \end{minipage}\hfill
    \begin{minipage}{0.5\textwidth}
        \centering
        \includegraphics[width=1\textwidth]{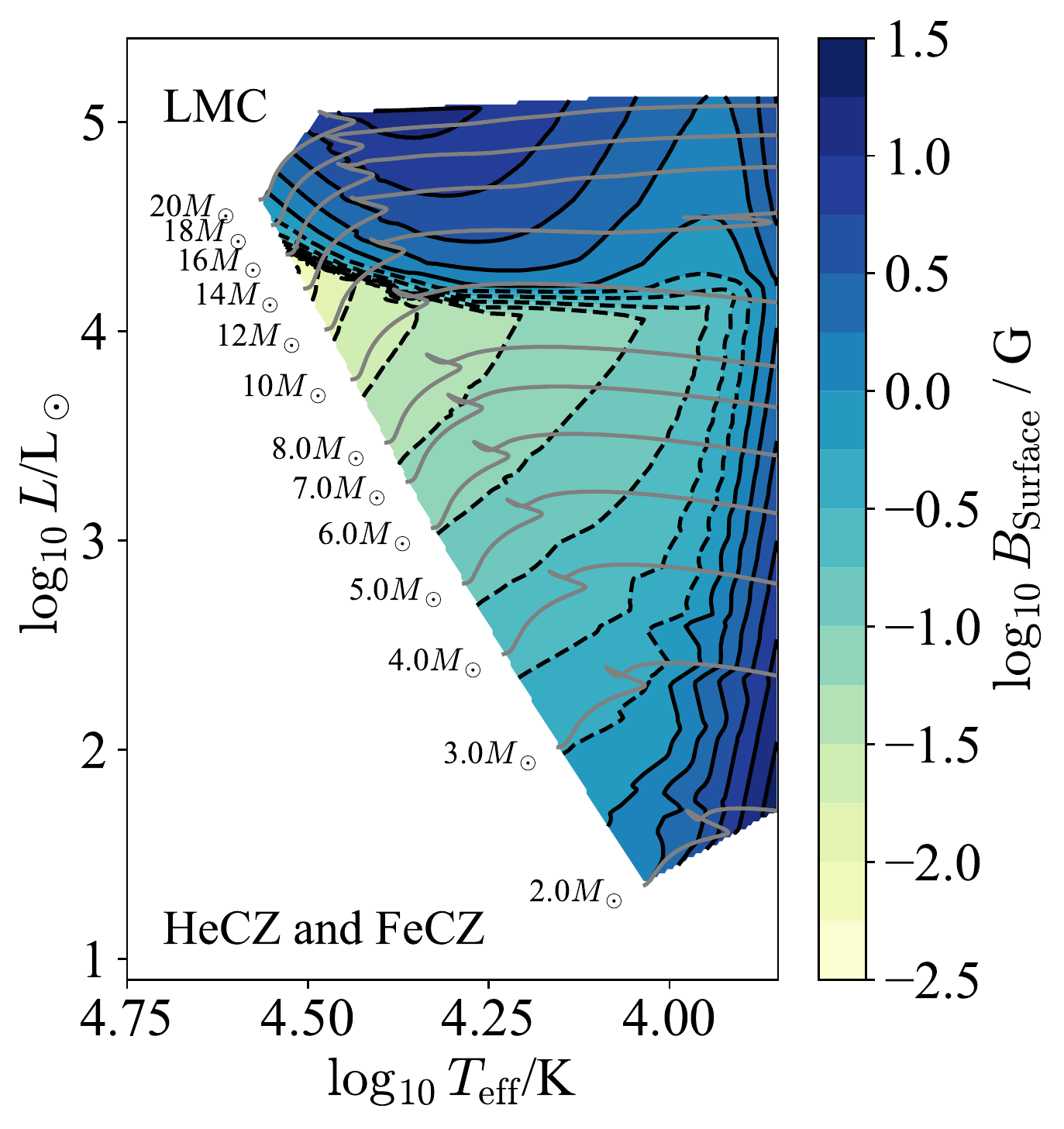} % second figure itself
    \end{minipage}
\caption{Same as Fig.~\ref{fig:MW_Bcrit}  but for stellar models ranging from $2-20 M_\odot$ with Z = 0.006.}\label{fig:LMC_Bcrit}
\end{figure}

\begin{figure}
    \centering
    \begin{minipage}{0.48\textwidth}
        \centering
        \includegraphics[width=1\textwidth]{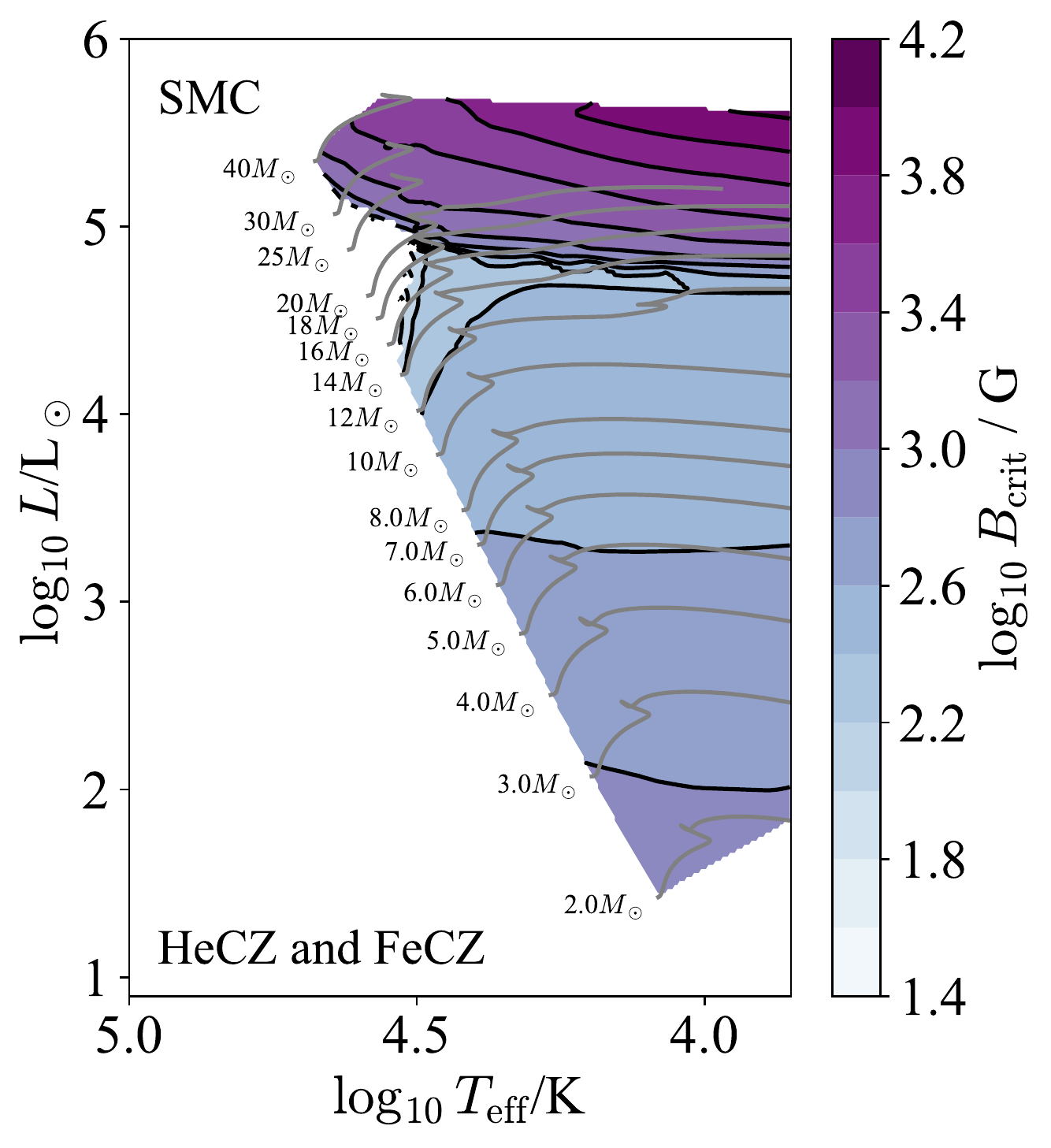} % first figure itself
    \end{minipage}\hfill
    \begin{minipage}{0.5\textwidth}
        \centering
        \includegraphics[width=1\textwidth]{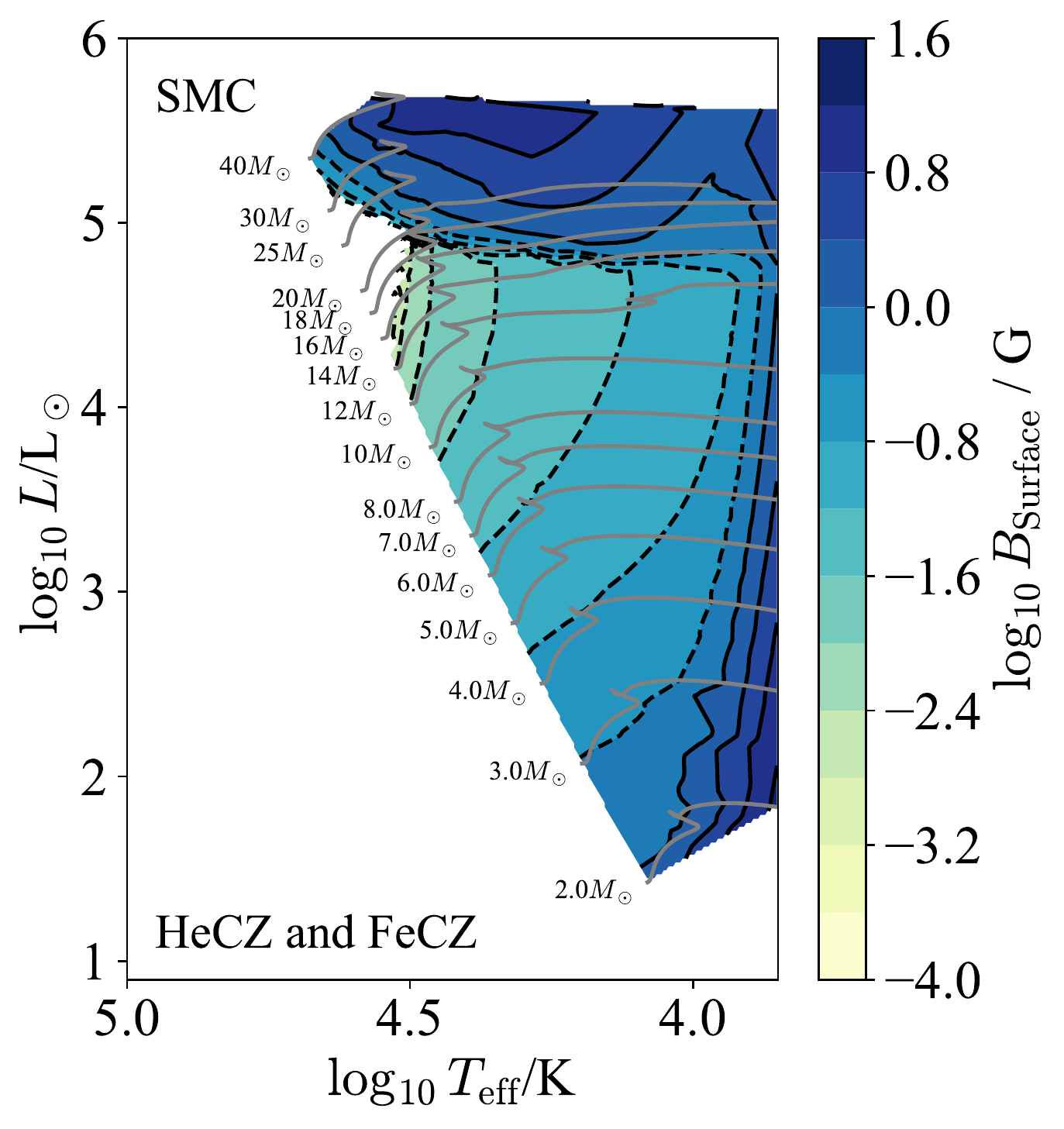} % second figure itself
    \end{minipage}
    \caption{Same as Fig.~\ref{fig:MW_Bcrit}  but for stellar models ranging from $2-40 M_\odot$ with Z = 0.002.}\label{fig:SMC_Bcrit}
\end{figure}

\bibliography{refs}

\begin{thebibliography}{}
\expandafter\ifx\csname natexlab\endcsname\relax\def\natexlab#1{#1}\fi
\providecommand{\url}[1]{\href{#1}{#1}}
\providecommand{\dodoi}[1]{doi:~\href{http://doi.org/#1}{\nolinkurl{#1}}}
\providecommand{\doeprint}[1]{\href{http://ascl.net/#1}{\nolinkurl{http://ascl.net/#1}}}
\providecommand{\doarXiv}[1]{\href{https://arxiv.org/abs/#1}{\nolinkurl{https://arxiv.org/abs/#1}}}

\bibitem[{{Alastuey} \& {Jancovici}(1978)}]{Alastuey1978}
{Alastuey}, A., \& {Jancovici}, B. 1978, \apj, 226, 1034,
  \dodoi{10.1086/156681}

\bibitem[{{Angulo} {et~al.}(1999){Angulo}, {Arnould}, {Rayet}, {Descouvemont},
  {Baye}, {Leclercq-Willain}, {Coc}, {Barhoumi}, {Aguer}, {Rolfs}, {Kunz},
  {Hammer}, {Mayer}, {Paradellis}, {Kossionides}, {Chronidou}, {Spyrou},
  {degl'Innocenti}, {Fiorentini}, {Ricci}, {Zavatarelli}, {Providencia},
  {Wolters}, {Soares}, {Grama}, {Rahighi}, {Shotter}, \& {Lamehi
  Rachti}}]{Angulo1999}
{Angulo}, C., {Arnould}, M., {Rayet}, M., {et~al.} 1999, Nuclear Physics A,
  656, 3, \dodoi{10.1016/S0375-9474(99)00030-5}

\bibitem[{{Aubert} {et~al.}(2017){Aubert}, {Gastine}, \&
  {Fournier}}]{2017JFM...813..558A}
{Aubert}, J., {Gastine}, T., \& {Fournier}, A. 2017, Journal of Fluid
  Mechanics, 813, 558, \dodoi{10.1017/jfm.2016.789}

\bibitem[{{Augustson} {et~al.}(2016){Augustson}, {Brun}, \&
  {Toomre}}]{2016ApJ...829...92A}
{Augustson}, K.~C., {Brun}, A.~S., \& {Toomre}, J. 2016, \apj, 829, 92,
  \dodoi{10.3847/0004-637X/829/2/92}

\bibitem[{{Auri{\`e}re} {et~al.}(2007){Auri{\`e}re}, {Wade}, {Silvester},
  {Ligni{\`e}res}, {Bagnulo}, {Bale}, {Dintrans}, {Donati}, {Folsom},
  {Gruberbauer}, {Hui Bon Hoa}, {Jeffers}, {Johnson}, {Landstreet},
  {L{\`e}bre}, {Lueftinger}, {Marsden}, {Mouillet}, {Naseri}, {Paletou},
  {Petit}, {Power}, {Rincon}, {Strasser}, \& {Toqu{\'e}}}]{2007A&A...475.1053A}
{Auri{\`e}re}, M., {Wade}, G.~A., {Silvester}, J., {et~al.} 2007, \aap, 475,
  1053, \dodoi{10.1051/0004-6361:20078189}

\bibitem[{{Auri{\`e}re} {et~al.}(2010){Auri{\`e}re}, {Wade}, {Ligni{\`e}res},
  {Hui-Bon-Hoa}, {Landstreet}, {Iliev}, {Donati}, {Petit}, {Roudier}, \&
  {Th{\'e}ado}}]{2010A&A...523A..40A}
{Auri{\`e}re}, M., {Wade}, G.~A., {Ligni{\`e}res}, F., {et~al.} 2010, \aap,
  523, A40, \dodoi{10.1051/0004-6361/201014848}

\bibitem[{{Avdeev} {et~al.}(1989){Avdeev}, {Dogel'}, \&
  {Dolgov}}]{1989ZhETF..96..885A}
{Avdeev}, E.~I., {Dogel'}, V.~A., \& {Dolgov}, O.~V. 1989, Zhurnal
  Eksperimentalnoi i Teoreticheskoi Fiziki, 96, 885

\bibitem[{{Babel}(1993)}]{1993ASPC...44..458B}
{Babel}, J. 1993, Astronomical Society of the Pacific Conference Series,
  Vol.~44, {Diffusion Models for Magnetic Ap-Stars}, ed. M.~M. {Dworetsky},
  F.~{Castelli}, \& R.~{Faraggiana}, 458

\bibitem[{{Balona}(2017)}]{2017MNRAS.467.1830B}
{Balona}, L.~A. 2017, \mnras, 467, 1830, \dodoi{10.1093/mnras/stx265}

\bibitem[{{Balona}(2019)}]{2019MNRAS.490.2112B}
---. 2019, \mnras, 490, 2112, \dodoi{10.1093/mnras/stz2808}

\bibitem[{Berdyugina(2005)}]{Berdyugina2005}
Berdyugina, S.~V. 2005, Living Reviews in Solar Physics, 2,
  \dodoi{10.12942/lrsp-2005-8}

\bibitem[{{Bouvier} {et~al.}(2007){Bouvier}, {Alencar}, {Harries},
  {Johns-Krull}, \& {Romanova}}]{2007prpl.conf..479B}
{Bouvier}, J., {Alencar}, S.~H.~P., {Harries}, T.~J., {Johns-Krull}, C.~M., \&
  {Romanova}, M.~M. 2007, in Protostars and Planets V, ed. B.~{Reipurth},
  D.~{Jewitt}, \& K.~{Keil}, 479.
\newblock \doarXiv{astro-ph/0603498}

\bibitem[{{Bowman} {et~al.}(2019){Bowman}, {Burssens}, {Pedersen}, {Johnston},
  {Aerts}, {Buysschaert}, {Michielsen}, {Tkachenko}, {Rogers}, {Edelmann},
  {Ratnasingam}, {Sim{\'o}n-D{\'\i}az}, {Castro}, {Moravveji}, {Pope}, {White},
  \& {De Cat}}]{2019NatAs...3..760B}
{Bowman}, D.~M., {Burssens}, S., {Pedersen}, M.~G., {et~al.} 2019, Nature
  Astronomy, 3, 760, \dodoi{10.1038/s41550-019-0768-1}

\bibitem[{{Braithwaite}(2009)}]{2009MNRAS.397..763B}
{Braithwaite}, J. 2009, \mnras, 397, 763,
  \dodoi{10.1111/j.1365-2966.2008.14034.x}

\bibitem[{{Braithwaite} \& {Cantiello}(2013)}]{2013MNRAS.428.2789B}
{Braithwaite}, J., \& {Cantiello}, M. 2013, \mnras, 428, 2789,
  \dodoi{10.1093/mnras/sts109}

\bibitem[{{Braithwaite} \& {Nordlund}(2006)}]{2006A&A...450.1077B}
{Braithwaite}, J., \& {Nordlund}, {\r{A}}. 2006, \aap, 450, 1077,
  \dodoi{10.1051/0004-6361:20041980}

\bibitem[{{Braithwaite} \& {Spruit}(2017)}]{2017RSOS....460271B}
{Braithwaite}, J., \& {Spruit}, H.~C. 2017, Royal Society Open Science, 4,
  160271, \dodoi{10.1098/rsos.160271}

\bibitem[{{Briquet} {et~al.}(2007){Briquet}, {Hubrig}, {De Cat}, {Aerts},
  {North}, \& {Sch{\"o}ller}}]{2007A&A...466..269B}
{Briquet}, M., {Hubrig}, S., {De Cat}, P., {et~al.} 2007, \aap, 466, 269,
  \dodoi{10.1051/0004-6361:20066940}

\bibitem[{{Brown} {et~al.}(2009){Brown}, {Browning}, {Brun}, {Miesch}, \&
  {Toomre}}]{2009ASPC..416..369B}
{Brown}, B.~P., {Browning}, M.~K., {Brun}, A.~S., {Miesch}, M.~S., \& {Toomre},
  J. 2009, Astronomical Society of the Pacific Conference Series, Vol. 416,
  {Dynamo Action and Wreaths of Magnetism in a Younger Sun}, ed. M.~{Dikpati},
  T.~{Arentoft}, I.~{Gonz{\'a}lez Hern{\'a}ndez}, C.~{Lindsey}, \& F.~{Hill},
  369

\bibitem[{{Buchler} \& {Yueh}(1976)}]{Buchler1976}
{Buchler}, J.~R., \& {Yueh}, W.~R. 1976, \apj, 210, 440, \dodoi{10.1086/154847}

\bibitem[{{Buysschaert} {et~al.}(2018){Buysschaert}, {Neiner}, {Martin},
  {Oksala}, {Aerts}, {Tkachenko}, {Alecian}, \& {the MiMeS
  Collaboration}}]{2018arXiv180805503B}
{Buysschaert}, B., {Neiner}, C., {Martin}, A.~J., {et~al.} 2018, arXiv
  e-prints, arXiv:1808.05503.
\newblock \doarXiv{1808.05503}

\bibitem[{{Cantiello} \& {Braithwaite}(2011)}]{2011A&A...534A.140C}
{Cantiello}, M., \& {Braithwaite}, J. 2011, \aap, 534, A140,
  \dodoi{10.1051/0004-6361/201117512}

\bibitem[{{Cantiello} \& {Braithwaite}(2019)}]{2019ApJ...883..106C}
---. 2019, \apj, 883, 106, \dodoi{10.3847/1538-4357/ab3924}

\bibitem[{{Cantiello} {et~al.}(2011){Cantiello}, {Braithwaite}, {Brandenburg},
  {Del Sordo}, {K{\"a}pyl{\"a}}, \& {Langer}}]{2011IAUS..272...32C}
{Cantiello}, M., {Braithwaite}, J., {Brandenburg}, A., {et~al.} 2011, in IAU
  Symposium, Vol. 272, Active OB Stars: Structure, Evolution, Mass Loss, and
  Critical Limits, ed. C.~{Neiner}, G.~{Wade}, G.~{Meynet}, \& G.~{Peters},
  32--37, \dodoi{10.1017/S174392131100994X}

\bibitem[{{Cantiello} {et~al.}(2009){Cantiello}, {Langer}, {Brott}, {de Koter},
  {Shore}, {Vink}, {Voegler}, {Lennon}, \& {Yoon}}]{2009A&A...499..279C}
{Cantiello}, M., {Langer}, N., {Brott}, I., {et~al.} 2009, \aap, 499, 279,
  \dodoi{10.1051/0004-6361/200911643}

\bibitem[{{Cassisi} {et~al.}(2007){Cassisi}, {Potekhin}, {Pietrinferni},
  {Catelan}, \& {Salaris}}]{Cassisi2007}
{Cassisi}, S., {Potekhin}, A.~Y., {Pietrinferni}, A., {Catelan}, M., \&
  {Salaris}, M. 2007, \apj, 661, 1094, \dodoi{10.1086/516819}

\bibitem[{{Castro} {et~al.}(2015){Castro}, {Fossati}, {Hubrig},
  {Sim{\'o}n-D{\'\i}az}, {Sch{\"o}ller}, {Ilyin}, {Carrol}, {Langer}, {Morel},
  {Schneider}, {Przybilla}, {Herrero}, {de Koter}, {Oskinova}, {Reisenegger},
  {Sana}, \& {BOB Collaboration}}]{2015A&A...581A..81C}
{Castro}, N., {Fossati}, L., {Hubrig}, S., {et~al.} 2015, \aap, 581, A81,
  \dodoi{10.1051/0004-6361/201425354}

\bibitem[{Christensen \& Aubert(2006)}]{doi:10.1111/j.1365-246X.2006.03009.x}
Christensen, U.~R., \& Aubert, J. 2006, Geophysical Journal International, 166,
  97, \dodoi{10.1111/j.1365-246X.2006.03009.x}

\bibitem[{{Chugunov} {et~al.}(2007){Chugunov}, {Dewitt}, \&
  {Yakovlev}}]{Chugunov2007}
{Chugunov}, A.~I., {Dewitt}, H.~E., \& {Yakovlev}, D.~G. 2007, \prd, 76,
  025028, \dodoi{10.1103/PhysRevD.76.025028}

\bibitem[{{Cowling}(1945)}]{1945MNRAS.105..166C}
{Cowling}, T.~G. 1945, \mnras, 105, 166, \dodoi{10.1093/mnras/105.3.166}

\bibitem[{{Currie} {et~al.}(2020){Currie}, {Barker}, {Lithwick}, \&
  {Browning}}]{2020MNRAS.493.5233C}
{Currie}, L.~K., {Barker}, A.~J., {Lithwick}, Y., \& {Browning}, M.~K. 2020,
  \mnras, 493, 5233, \dodoi{10.1093/mnras/staa372}

\bibitem[{{Cyburt} {et~al.}(2010){Cyburt}, {Amthor}, {Ferguson}, {Meisel},
  {Smith}, {Warren}, {Heger}, {Hoffman}, {Rauscher}, {Sakharuk}, {Schatz},
  {Thielemann}, \& {Wiescher}}]{Cyburt2010}
{Cyburt}, R.~H., {Amthor}, A.~M., {Ferguson}, R., {et~al.} 2010, \apjs, 189,
  240, \dodoi{10.1088/0067-0049/189/1/240}

\bibitem[{{Dewitt} {et~al.}(1973){Dewitt}, {Graboske}, \&
  {Cooper}}]{Dewitt1973}
{Dewitt}, H.~E., {Graboske}, H.~C., \& {Cooper}, M.~S. 1973, \apj, 181, 439,
  \dodoi{10.1086/152061}

\bibitem[{{Duez} \& {Mathis}(2010)}]{2010A&A...517A..58D}
{Duez}, V., \& {Mathis}, S. 2010, \aap, 517, A58,
  \dodoi{10.1051/0004-6361/200913496}

\bibitem[{{Ferguson} {et~al.}(2005){Ferguson}, {Alexander}, {Allard}, {Barman},
  {Bodnarik}, {Hauschildt}, {Heffner-Wong}, \& {Tamanai}}]{Ferguson2005}
{Ferguson}, J.~W., {Alexander}, D.~R., {Allard}, F., {et~al.} 2005, \apj, 623,
  585, \dodoi{10.1086/428642}

\bibitem[{{Fossati} {et~al.}(2015{\natexlab{a}}){Fossati}, {Castro}, {Morel},
  {Langer}, {Briquet}, {Carroll}, {Hubrig}, {Nieva}, {Oskinova}, {Przybilla},
  {Schneider}, {Sch{\"o}ller}, {Sim{\'o}n-D{\'\i}az}, {Ilyin}, {de Koter},
  {Reisenegger}, \& {Sana}}]{2015A&A...574A..20F}
{Fossati}, L., {Castro}, N., {Morel}, T., {et~al.} 2015{\natexlab{a}}, \aap,
  574, A20, \dodoi{10.1051/0004-6361/201424986}

\bibitem[{{Fossati} {et~al.}(2015{\natexlab{b}}){Fossati}, {Castro},
  {Sch{\"o}ller}, {Hubrig}, {Langer}, {Morel}, {Briquet}, {Herrero},
  {Przybilla}, {Sana}, {Schneider}, {de Koter}, \& {BOB
  Collaboration}}]{2015A&A...582A..45F}
{Fossati}, L., {Castro}, N., {Sch{\"o}ller}, M., {et~al.} 2015{\natexlab{b}},
  \aap, 582, A45, \dodoi{10.1051/0004-6361/201526725}

\bibitem[{{Fossati} {et~al.}(2016){Fossati}, {Schneider}, {Castro}, {Langer},
  {Sim{\'o}n-D{\'\i}az}, {M{\"u}ller}, {de Koter}, {Morel}, {Petit}, {Sana}, \&
  {Wade}}]{2016A&A...592A..84F}
{Fossati}, L., {Schneider}, F.~R.~N., {Castro}, N., {et~al.} 2016, \aap, 592,
  A84, \dodoi{10.1051/0004-6361/201628259}

\bibitem[{{Fuller} {et~al.}(1985){Fuller}, {Fowler}, \& {Newman}}]{Fuller1985}
{Fuller}, G.~M., {Fowler}, W.~A., \& {Newman}, M.~J. 1985, \apj, 293, 1,
  \dodoi{10.1086/163208}

\bibitem[{{Fuller} {et~al.}(2019){Fuller}, {Piro}, \&
  {Jermyn}}]{2019MNRAS.485.3661F}
{Fuller}, J., {Piro}, A.~L., \& {Jermyn}, A.~S. 2019, \mnras, 485, 3661,
  \dodoi{10.1093/mnras/stz514}

\bibitem[{{Gough} \& {Tayler}(1966)}]{1966MNRAS.133...85G}
{Gough}, D.~O., \& {Tayler}, R.~J. 1966, \mnras, 133, 85,
  \dodoi{10.1093/mnras/133.1.85}

\bibitem[{{Grassitelli} {et~al.}(2015){Grassitelli}, {Fossati},
  {Sim{\'o}n-Di{\'a}z}, {Langer}, {Castro}, \& {Sanyal}}]{2015ApJ...808L..31G}
{Grassitelli}, L., {Fossati}, L., {Sim{\'o}n-Di{\'a}z}, S., {et~al.} 2015,
  \apjl, 808, L31, \dodoi{10.1088/2041-8205/808/1/L31}

\bibitem[{{Grevesse} \& {Sauval}(1998)}]{1998SSRv...85..161G}
{Grevesse}, N., \& {Sauval}, A.~J. 1998, \ssr, 85, 161,
  \dodoi{10.1023/A:1005161325181}

\bibitem[{{Grunhut} {et~al.}(2017){Grunhut}, {Wade}, {Neiner}, {Oksala},
  {Petit}, {Alecian}, {Bohlender}, {Bouret}, {Henrichs}, {Hussain},
  {Kochukhov}, \& {MiMeS Collaboration}}]{2017MNRAS.465.2432G}
{Grunhut}, J.~H., {Wade}, G.~A., {Neiner}, C., {et~al.} 2017, \mnras, 465,
  2432, \dodoi{10.1093/mnras/stw2743}

\bibitem[{{Harrington} \& {Garaud}(2019)}]{2019ApJ...870L...5H}
{Harrington}, P.~Z., \& {Garaud}, P. 2019, \apjl, 870, L5,
  \dodoi{10.3847/2041-8213/aaf812}

\bibitem[{{Huang} \& {Gies}(2008)}]{2008ApJ...683.1045H}
{Huang}, W., \& {Gies}, D.~R. 2008, \apj, 683, 1045, \dodoi{10.1086/590106}

\bibitem[{{Iglesias} \& {Rogers}(1993)}]{Iglesias1993}
{Iglesias}, C.~A., \& {Rogers}, F.~J. 1993, \apj, 412, 752,
  \dodoi{10.1086/172958}

\bibitem[{{Iglesias} \& {Rogers}(1996)}]{Iglesias1996}
---. 1996, \apj, 464, 943, \dodoi{10.1086/177381}

\bibitem[{{Itoh} {et~al.}(1996){Itoh}, {Hayashi}, {Nishikawa}, \&
  {Kohyama}}]{Itoh1996}
{Itoh}, N., {Hayashi}, H., {Nishikawa}, A., \& {Kohyama}, Y. 1996, \apjs, 102,
  411, \dodoi{10.1086/192264}

\bibitem[{{Itoh} {et~al.}(1979){Itoh}, {Totsuji}, {Ichimaru}, \&
  {Dewitt}}]{Itoh1979}
{Itoh}, N., {Totsuji}, H., {Ichimaru}, S., \& {Dewitt}, H.~E. 1979, \apj, 234,
  1079, \dodoi{10.1086/157590}

\bibitem[{Jermyn \& Cantiello(2020)}]{data}
Jermyn, A., \& Cantiello, M. 2020, {Data for "The Origin of the Bimodal
  Distribution of Magnetic Fields in Early-type Stars"}, MESA 11701,  Zenodo,
  \dodoi{10.5281/zenodo.3891940}

\bibitem[{{Jiang} {et~al.}(2015){Jiang}, {Cantiello}, {Bildsten}, {Quataert},
  \& {Blaes}}]{2015ApJ...813...74J}
{Jiang}, Y.-F., {Cantiello}, M., {Bildsten}, L., {Quataert}, E., \& {Blaes}, O.
  2015, \apj, 813, 74, \dodoi{10.1088/0004-637X/813/1/74}

\bibitem[{{Jiang} {et~al.}(2018){Jiang}, {Cantiello}, {Bildsten}, {Quataert},
  {Blaes}, \& {Stone}}]{2018Natur.561..498J}
{Jiang}, Y.-F., {Cantiello}, M., {Bildsten}, L., {et~al.} 2018, \nat, 561, 498,
  \dodoi{10.1038/s41586-018-0525-0}

\bibitem[{{Kholtygin} {et~al.}(2010){Kholtygin}, {Fabrika}, {Drake}, {Bychkov},
  {Bychkova}, {Chountonov}, {Burlakova}, \& {Valyavin}}]{2010AstL...36..370K}
{Kholtygin}, A.~F., {Fabrika}, S.~N., {Drake}, N.~A., {et~al.} 2010, Astronomy
  Letters, 36, 370, \dodoi{10.1134/S1063773710050087}

\bibitem[{{Landstreet} {et~al.}(1989){Landstreet}, {Barker}, {Bohlender}, \&
  {Jewison}}]{1989ApJ...344..876L}
{Landstreet}, J.~D., {Barker}, P.~K., {Bohlender}, D.~A., \& {Jewison}, M.~S.
  1989, \apj, 344, 876, \dodoi{10.1086/167855}

\bibitem[{{Langanke} \& {Mart{\'{\i}}nez-Pinedo}(2000)}]{Langanke2000}
{Langanke}, K., \& {Mart{\'{\i}}nez-Pinedo}, G. 2000, Nuclear Physics A, 673,
  481, \dodoi{10.1016/S0375-9474(00)00131-7}

\bibitem[{{Lecoanet} {et~al.}(2019){Lecoanet}, {Cantiello}, {Quataert},
  {Couston}, {Burns}, {Pope}, {Jermyn}, {Favier}, \& {Le
  Bars}}]{2019ApJ...886L..15L}
{Lecoanet}, D., {Cantiello}, M., {Quataert}, E., {et~al.} 2019, \apjl, 886,
  L15, \dodoi{10.3847/2041-8213/ab5446}

\bibitem[{{Ligni{\`e}res} {et~al.}(2014){Ligni{\`e}res}, {Petit},
  {Auri{\`e}re}, {Wade}, \& {B{\"o}hm}}]{2014IAUS..302..338L}
{Ligni{\`e}res}, F., {Petit}, P., {Auri{\`e}re}, M., {Wade}, G.~A., \&
  {B{\"o}hm}, T. 2014, in IAU Symposium, Vol. 302, Magnetic Fields throughout
  Stellar Evolution, ed. P.~{Petit}, M.~{Jardine}, \& H.~C. {Spruit}, 338--347,
  \dodoi{10.1017/S1743921314002440}

\bibitem[{{MacDonald} \& {Mullan}(2009)}]{2009ApJ...700..387M}
{MacDonald}, J., \& {Mullan}, D.~J. 2009, \apj, 700, 387,
  \dodoi{10.1088/0004-637X/700/1/387}

\bibitem[{{MacDonald} \& {Petit}(2019)}]{2019MNRAS.487.3904M}
{MacDonald}, J., \& {Petit}, V. 2019, \mnras, 487, 3904,
  \dodoi{10.1093/mnras/stz1545}

\bibitem[{{Medvedev} {et~al.}(2018){Medvedev}, {Kholtygin}, {Hubrig},
  {Fabrika}, {Valyavin}, {Sch{\"o}ller}, \& {Tsiopa}}]{2018CoSka..48..223M}
{Medvedev}, A.~S., {Kholtygin}, A.~F., {Hubrig}, S., {et~al.} 2018,
  Contributions of the Astronomical Observatory Skalnate Pleso, 48, 223

\bibitem[{{Moreno-Insertis} \& {Spruit}(1989)}]{1989ApJ...342.1158M}
{Moreno-Insertis}, F., \& {Spruit}, H.~C. 1989, \apj, 342, 1158,
  \dodoi{10.1086/167673}

\bibitem[{{Moss}(1987)}]{1987MNRAS.224.1019M}
{Moss}, D. 1987, \mnras, 224, 1019, \dodoi{10.1093/mnras/224.4.1019}

\bibitem[{{Oda} {et~al.}(1994){Oda}, {Hino}, {Muto}, {Takahara}, \&
  {Sato}}]{Oda1994}
{Oda}, T., {Hino}, M., {Muto}, K., {Takahara}, M., \& {Sato}, K. 1994, Atomic
  Data and Nuclear Data Tables, 56, 231, \dodoi{10.1006/adnd.1994.1007}

\bibitem[{{Paxton} {et~al.}(2011){Paxton}, {Bildsten}, {Dotter}, {Herwig},
  {Lesaffre}, \& {Timmes}}]{Paxton2011}
{Paxton}, B., {Bildsten}, L., {Dotter}, A., {et~al.} 2011, \apjs, 192, 3,
  \dodoi{10.1088/0067-0049/192/1/3}

\bibitem[{{Paxton} {et~al.}(2013){Paxton}, {Cantiello}, {Arras}, {Bildsten},
  {Brown}, {Dotter}, {Mankovich}, {Montgomery}, {Stello}, {Timmes}, \&
  {Townsend}}]{Paxton2013}
{Paxton}, B., {Cantiello}, M., {Arras}, P., {et~al.} 2013, \apjs, 208, 4,
  \dodoi{10.1088/0067-0049/208/1/4}

\bibitem[{{Paxton} {et~al.}(2015){Paxton}, {Marchant}, {Schwab}, {Bauer},
  {Bildsten}, {Cantiello}, {Dessart}, {Farmer}, {Hu}, {Langer}, {Townsend},
  {Townsley}, \& {Timmes}}]{Paxton2015}
{Paxton}, B., {Marchant}, P., {Schwab}, J., {et~al.} 2015, \apjs, 220, 15,
  \dodoi{10.1088/0067-0049/220/1/15}

\bibitem[{{Paxton} {et~al.}(2018){Paxton}, {Schwab}, {Bauer}, {Bildsten},
  {Blinnikov}, {Duffell}, {Farmer}, {Goldberg}, {Marchant}, {Sorokina},
  {Thoul}, {Townsend}, \& {Timmes}}]{Paxton2018}
{Paxton}, B., {Schwab}, J., {Bauer}, E.~B., {et~al.} 2018, \apjs, 234, 34,
  \dodoi{10.3847/1538-4365/aaa5a8}

\bibitem[{{Paxton} {et~al.}(2019){Paxton}, {Smolec}, {Schwab}, {Gautschy},
  {Bildsten}, {Cantiello}, {Dotter}, {Farmer}, {Goldberg}, {Jermyn}, {Kanbur},
  {Marchant}, {Thoul}, {Townsend}, {Wolf}, {Zhang}, \& {Timmes}}]{Paxton2019}
{Paxton}, B., {Smolec}, R., {Schwab}, J., {et~al.} 2019, \apjs, 243, 10,
  \dodoi{10.3847/1538-4365/ab2241}

\bibitem[{{Petit} {et~al.}(2010){Petit}, {Ligni{\`e}res}, {Wade},
  {Auri{\`e}re}, {B{\"o}hm}, {Bagnulo}, {Dintrans}, {Fumel}, {Grunhut},
  {Lanoux}, {Morgenthaler}, \& {Van Grootel}}]{2010A&A...523A..41P}
{Petit}, P., {Ligni{\`e}res}, F., {Wade}, G.~A., {et~al.} 2010, \aap, 523, A41,
  \dodoi{10.1051/0004-6361/201015307}

\bibitem[{{Petit} {et~al.}(2011){Petit}, {Ligni{\`e}res}, {Auri{\`e}re},
  {Wade}, {Alina}, {Ballot}, {B{\"o}hm}, {Jouve}, {Oza}, {Paletou}, \&
  {Th{\'e}ado}}]{2011A&A...532L..13P}
{Petit}, P., {Ligni{\`e}res}, F., {Auri{\`e}re}, M., {et~al.} 2011, \aap, 532,
  L13, \dodoi{10.1051/0004-6361/201117573}

\bibitem[{{Pols} {et~al.}(1995){Pols}, {Tout}, {Eggleton}, \& {Han}}]{Pols1995}
{Pols}, O.~R., {Tout}, C.~A., {Eggleton}, P.~P., \& {Han}, Z. 1995, \mnras,
  274, 964, \dodoi{10.1093/mnras/274.3.964}

\bibitem[{{Potekhin} \& {Chabrier}(2010)}]{Potekhin2010}
{Potekhin}, A.~Y., \& {Chabrier}, G. 2010, Contributions to Plasma Physics, 50,
  82, \dodoi{10.1002/ctpp.201010017}

\bibitem[{{Power} {et~al.}(2007){Power}, {Wade}, {Hanes}, {Aurier}, \&
  {Silvester}}]{2007pms..conf...89P}
{Power}, J., {Wade}, G.~A., {Hanes}, D.~A., {Aurier}, M., \& {Silvester}, J.
  2007, in Physics of Magnetic Stars, ed. I.~I. {Romanyuk}, D.~O.
  {Kudryavtsev}, O.~M. {Neizvestnaya}, \& V.~M. {Shapoval}, 89--97.
\newblock \doarXiv{astro-ph/0612557}

\bibitem[{{Rogers} \& {Nayfonov}(2002)}]{Rogers2002}
{Rogers}, F.~J., \& {Nayfonov}, A. 2002, \apj, 576, 1064,
  \dodoi{10.1086/341894}

\bibitem[{{Salpeter}(1954)}]{Salpeter1954}
{Salpeter}, E.~E. 1954, Australian Journal of Physics, 7, 373,
  \dodoi{10.1071/PH540373}

\bibitem[{{Saumon} {et~al.}(1995){Saumon}, {Chabrier}, \& {van
  Horn}}]{Saumon1995}
{Saumon}, D., {Chabrier}, G., \& {van Horn}, H.~M. 1995, \apjs, 99, 713,
  \dodoi{10.1086/192204}

\bibitem[{Schneider {et~al.}(2019)Schneider, Ohlmann, Podsiadlowski, R\"{o}pke,
  Balbus, Pakmor, \& Springel}]{Schneider2019}
Schneider, F. R.~N., Ohlmann, S.~T., Podsiadlowski, P., {et~al.} 2019, Nature,
  574, 211, \dodoi{10.1038/s41586-019-1621-5}

\bibitem[{{Sikora} {et~al.}(2019){Sikora}, {Wade}, {Power}, \&
  {Neiner}}]{2019MNRAS.483.3127S}
{Sikora}, J., {Wade}, G.~A., {Power}, J., \& {Neiner}, C. 2019, \mnras, 483,
  3127, \dodoi{10.1093/mnras/sty2895}

\bibitem[{{Sim{\'o}n-D{\'\i}az} {et~al.}(2017){Sim{\'o}n-D{\'\i}az}, {Godart},
  {Castro}, {Herrero}, {Aerts}, {Puls}, {Telting}, \&
  {Grassitelli}}]{2017A&A...597A..22S}
{Sim{\'o}n-D{\'\i}az}, S., {Godart}, M., {Castro}, N., {et~al.} 2017, \aap,
  597, A22, \dodoi{10.1051/0004-6361/201628541}

\bibitem[{{Sim{\'o}n-D{\'\i}az} \& {Herrero}(2014)}]{2014A&A...562A.135S}
{Sim{\'o}n-D{\'\i}az}, S., \& {Herrero}, A. 2014, \aap, 562, A135,
  \dodoi{10.1051/0004-6361/201322758}

\bibitem[{{Spruit}(2002)}]{2002A&A...381..923S}
{Spruit}, H.~C. 2002, \aap, 381, 923, \dodoi{10.1051/0004-6361:20011465}

\bibitem[{{Stevenson}(1982)}]{1982GApFD..21..113S}
{Stevenson}, D.~J. 1982, Geophysical and Astrophysical Fluid Dynamics, 21, 113,
  \dodoi{10.1080/03091928208209008}

\bibitem[{Sundqvist {et~al.}(2013)Sundqvist, Petit, Owocki, Wade, Puls, \&
  Collaboration}]{10.1093/mnras/stt921}
Sundqvist, J.~O., Petit, V., Owocki, S.~P., {et~al.} 2013, Monthly Notices of
  the Royal Astronomical Society, 433, 2497, \dodoi{10.1093/mnras/stt921}

\bibitem[{{Sundqvist} {et~al.}(2013){Sundqvist}, {Petit}, {Owocki}, {Wade},
  {Puls}, \& {MiMeS Collaboration}}]{2013MNRAS.433.2497S}
{Sundqvist}, J.~O., {Petit}, V., {Owocki}, S.~P., {et~al.} 2013, \mnras, 433,
  2497, \dodoi{10.1093/mnras/stt921}

\bibitem[{{Timmes} \& {Swesty}(2000)}]{Timmes2000}
{Timmes}, F.~X., \& {Swesty}, F.~D. 2000, \apjs, 126, 501,
  \dodoi{10.1086/313304}

\bibitem[{{Trust} {et~al.}(2020){Trust}, {Jurua}, {De Cat}, \&
  {Joshi}}]{2020MNRAS.492.3143T}
{Trust}, O., {Jurua}, E., {De Cat}, P., \& {Joshi}, S. 2020, \mnras, 492, 3143,
  \dodoi{10.1093/mnras/stz3623}

\bibitem[{{ud-Doula} {et~al.}(2009){ud-Doula}, {Owocki}, \&
  {Townsend}}]{2009IAUS..259..423U}
{ud-Doula}, A., {Owocki}, S.~P., \& {Townsend}, R. H.~D. 2009, in IAU
  Symposium, Vol. 259, Cosmic Magnetic Fields: From Planets, to Stars and
  Galaxies, ed. K.~G. {Strassmeier}, A.~G. {Kosovichev}, \& J.~E. {Beckman},
  423--424, \dodoi{10.1017/S174392130903097X}

\bibitem[{{Weber} \& {Davis}(1967)}]{1967ApJ...148..217W}
{Weber}, E.~J., \& {Davis}, Leverett, J. 1967, \apj, 148, 217,
  \dodoi{10.1086/149138}

\bibitem[{{Yusof} {et~al.}(2013){Yusof}, {Hirschi}, {Meynet}, {Crowther},
  {Ekstr{\"o}m}, {Frischknecht}, {Georgy}, {Abu Kassim}, \&
  {Schnurr}}]{2013MNRAS.433.1114Y}
{Yusof}, N., {Hirschi}, R., {Meynet}, G., {et~al.} 2013, \mnras, 433, 1114,
  \dodoi{10.1093/mnras/stt794}

\bibitem[{Zeldovich(1957)}]{Zeldovich}
Zeldovich, Y.~B. 1957, Soviet Physics Journal of Experimental and Theoretical
  Physics, 460

\bibitem[{{Zorec} \& {Royer}(2012)}]{2012A&A...537A.120Z}
{Zorec}, J., \& {Royer}, F. 2012, \aap, 537, A120,
  \dodoi{10.1051/0004-6361/201117691}

\end{thebibliography}
\bibliographystyle{aasjournal}

\end{document}